\documentclass[useAMS,usenatbib]{mn2e}
\usepackage{natbib}
\usepackage{amsmath}
\usepackage{url}
\usepackage{longtable}
\usepackage{aas_macros}
\usepackage{amssymb}
\usepackage{graphicx}
\usepackage{deluxetable}

\newcommand{\jhks}{\protect\hbox{$J\!H\!K_{s}$} }

\newcommand{\about}{$\sim\!\!$~}
\newcommand{\kms}{\,km\,s$^{-1}$}

\def\lsim{\hbox{\rlap{\raise 0.425ex\hbox{$<$}}\lower 0.65ex\hbox{$\sim$}}}
\def\gsim{\hbox{\rlap{\raise 0.425ex\hbox{$>$}}\lower 0.65ex\hbox{$\sim$}}}

\def\arcsec{\hbox{$^{\prime\prime}$}}

\newcommand\ion[2]{#1$\,${\small{#2}}\relax}

\title[SN~2014J]{Extensive {\it HST} Ultraviolet Spectra and
  Multi-wavelength Observations of SN~2014J in M82 Indicate Reddening
  and Circumstellar Scattering by Typical Dust}

\def\illast{1}
\def\illphys{2}
\def\berk{3}
\def\rut{4}
\def\car{5}
\def\ttech{6}
\def\cfa{7}
\def\ipmu{8}
\def\mpia{9}
\def\stock{10}
\def\ut{11}
\def\uva{12}
\def\heid{13}
\def\ab{14}
\def\mil{15}
\def\wurz{16}
\def\strom{17}
\def\aar{18}

\author[Foley et~al.]{Ryan~J.~Foley$^{\illast,\illphys}$\thanks{E-mail:rfoley@illinois.edu},
O.~D.~Fox$^{\berk}$, %
C.~McCully$^{\rut}$, %
M.~M.~Phillips$^{\car}$, %
D.~J.~Sand$^{\ttech}$, %
W.~Zheng$^{\berk}$,
\newauthor
P.~Challis$^{\cfa}$, %
A.~V.~Filippenko$^{\berk}$, %
G.~Folatelli$^{\ipmu}$,
W.~Hillebrandt$^{\mpia}$, %
E.~Y.~Hsiao$^{\car}$, %
\newauthor
S.~W.~Jha$^{\rut}$, %
R.~P.~Kirshner$^{\cfa}$, %
M.~Kromer$^{\stock}$, %
G.~H.~Marion$^{\ut}$, %
M.~Nelson$^{\uva}$, %
\newauthor
R.~Pakmor$^{\heid}$, %
G.~Pignata$^{\ab,\mil}$, %
F.~K.~R\"{o}pke$^{\wurz}$, %
I.~R.~Seitenzahl$^{\strom}$, %
J.~M.~Silverman$^{\ut}$, %
\newauthor
M.~Skrutskie$^{\uva}$, %
M.~D.~Stritzinger$^{\aar}$\\
$^{\illast}$Astronomy Department, University of Illinois at Urbana--Champaign, 1002 W.\ Green Street, Urbana, IL 61801, USA\\
$^{\illphys}$Department of Physics, University of Illinois Urbana--Champaign, 1110 W.\ Green Street, Urbana, IL 61801, USA\\
$^{\berk}$Department of Astronomy, University of California, Berkeley, CA 94720-3411, USA\\
$^{\rut}$Department of Physics and Astronomy, Rutgers, the State University of New Jersey, 136 Frelinghuysen Road, Piscataway,\\ NJ 08854, USA\\
$^{\car}$Carnegie Observatories, Las Campanas Observatory, La Serena, Chile\\
$^{\ttech}$Physics Department, Texas Tech University, Lubbock, TX 79409, USA\\
$^{\cfa}$Harvard-Smithsonian Center for Astrophysics, 60 Garden Street, Cambridge, MA 02138, USA\\
$^{\ipmu}$Kavli Institute for the Physics and Mathematics of the Universe, Todai Institutes for Advanced Study, the University of Tokyo,\\ Kashiwa, Japan 277-8583 (Kavli IPMU, WPI)\\
$^{\mpia}$Max-Planck-Institut f\"{u}r Astrophysik, Karl-Schwarzschild-Strasse 1, D-85748 Garching bei M\"{u}nchen, Germany\\
$^{\stock}$The Oskar Klein Centre \& Department of Astronomy, Stockholm University, AlbaNova, SE-106 91 Stockholm, Sweden\\
$^{\ut}$Department of Astronomy, University of Texas, Austin, TX 78712-0259, USA\\
$^{\uva}$Department of Astronomy, University of Virginia, Charlottesville, VA 22904, USA\\
$^{\heid}$Heidelberger Institut f\"{u}r Theoretische Studien, Schloss-Wolfsbrunnenweg 35, D-69118 Heidelberg, Germany\\
$^{\ab}$Departamento de Ciencias Fisicas, Universidad Andres Bello, Avda.\ Republica 252, Santiago, Chile\\
$^{\mil}$Millennium Institute of Astrophysics, Avda.\ Republica 252, Santiago, Chile\\
$^{\wurz}$Institut f{\"u}r Theoretische Physik und Astrophysik, Universit{\"a}t W{\"u}rzburg, Emil-Fischer-Str.\ 31, D-97074 W{\"u}rzburg, Germany\\
$^{\strom}$Research School of Astronomy and Astrophysics, Mount Stromlo Observatory, Cotter Road, Weston Creek, ACT 2611, Australia\\
$^{\aar}$Department of Physics and Astronomy, Aarhus University, Ny Munkegade, DK-8000 Aarhus C, Denmark}

\begin{document}

\date{Accepted  . Received   ; in original form  }
\pagerange{\pageref{firstpage}--\pageref{lastpage}} \pubyear{2014}
\maketitle
\label{firstpage}

\begin{abstract}
  SN~2014J in M82 is the closest detected Type Ia supernova (SN~Ia) in
  at least 28 years and perhaps in 410 years.  Despite its small
  distance of 3.3~Mpc, SN~2014J is surprisingly faint, peaking at $V =
  10.6$~mag, and assuming a typical SN~Ia luminosity, we infer an
  observed visual extinction of $A_{V} = 2.0 \pm 0.1$~mag.  But this
  picture, with $R_{V} = 1.6 \pm 0.2$, is too simple to account for
  all observations.  We combine 10 epochs (spanning a month) of {\it
    HST}/STIS ultraviolet through near-infrared spectroscopy with {\it
    HST}/WFC3, KAIT, and FanCam photometry from the optical to the
  infrared and 9 epochs of high-resolution TRES spectroscopy to
  investigate the sources of extinction and reddening for SN~2014J.
  We argue that the wide range of observed properties for SN~2014J is
  caused by a combination of dust reddening, likely originating in the
  interstellar medium of M82, and scattering off circumstellar
  material.  For this model, roughly half of the extinction is caused
  by reddening from typical dust ($E(B-V) = 0.45$~mag and $R_{V} =
  2.6$) and roughly half by scattering off LMC-like dust in the
  circumstellar environment of SN~2014J.
\end{abstract}



\defcitealias{Fitzpatrick99}{F99}
\defcitealias{Cardelli89}{CCM}
\defcitealias{Goobar08}{G08}

\section{Introduction}\label{s:intro}

Type Ia supernovae (SNe~Ia) are rare enough that nearby SNe~Ia ($D <
5$~Mpc) are discovered only about once a decade.  But these rare
events, which provide unique and extremely high-quality data and can
be monitored for years, can lead to large jumps in our understanding
of SN physics.

Observations of SNe~Ia were used to discover the accelerating
expansion of the Universe \citep{Riess98:Lambda, Perlmutter99}, and
they continue to be one of our best cosmological probes
\citep[e.g.,][]{Rest13, Betoule14}.

One of the largest potential systematic uncertainties for using SNe~Ia
as cosmological probes is the poorly constrained and potentially
peculiar properties of absorbing dust along the line of sight to
SNe~Ia \citep[e.g.,][]{Scolnic14:ps1, Scolnic14:col}.  Understanding
these dust properties, often simplified as a single reddening-law
parameter, $R_{V}$, is critical for cosmological measurements since
the distance modulus measured to a SN~Ia is
\begin{equation}
  \mu = m_{V} - M_{V} + E(B-V) \cdot R_{V},
\end{equation}
where $m_{V}$ and $M_{V}$ are the apparent and absolute magnitudes of
a SN in the $V$ band, $E(B-V)$ is the reddening, and $R_{V}$ is the
ratio of the total-to-selective extinction.  Similar equations exist
for all bands.

Several different methods indicate that the dust reddening SNe~Ia has
$R_{V} < 2$, which is significantly below the average value --- and
below nearly the entire population --- for Milky Way lines of sight
\citep[e.g.,][]{Fitzpatrick07}.  SN reddening is measured by comparing
observed SN colours to a zero-reddening locus \citep[e.g.,][]{Riess96,
  Phillips99}.  Various methods of examining large samples of SNe~Ia
have resulted in $R_{V} < 2$ \citep[e.g.,][]{Nobili05, Guy05,
  Hicken09:de, Folatelli10, Burns11}.  It is now understood that at
least part of the reason for the low values of $R_{V}$ from large
samples is because of poor assumptions about the intrinsic colour
distribution of SNe~Ia \citep[e.g.,][]{Foley11:vel, Mandel11};
correcting for this effect can increase the best-fitting $R_{V}$ from
1.6 to 2.5 for a large sample of SNe~Ia \citep{Foley11:vel}.  However,
there are several examples of highly reddened SNe~Ia, where $R_{V}$
can be measured directly, that still have $R_{V} < 2$
\citep[e.g.,][]{Elias-Rosa06, Elias-Rosa08, Krisciunas06, Wang08:06x}.

As an alternative to the Milky Way having peculiar dust,
\citet{Wang05} suggested that circumstellar dust scattering will
naturally lead to low values of $R_{V}$.  Presumably SNe~Ia with
higher reddening are more likely to have additional, circumstellar
dust, which can potentially both scatter and redden, perhaps resolving
the low values of $R_{V}$ for the most highly reddened SNe~Ia.
\citet{Patat06} further investigated this possibility providing
several predictions for observations.  \citet[hereafter,
\citetalias{Goobar08}]{Goobar08} quantified this effect, producing
simulations of how the scattering will affect a SN spectral energy
distribution (SED) and the inferred $R_{V}$.

Adding further mystery to the situation are correlations between SN
observables and gas/dust properties.  Specifically, highly reddened
SNe~Ia with high-velocity ejecta tend to have lower values for $R_{V}$
relative to their equally reddened, low ejecta velocity counterparts
\citep{Wang09:2pop}.  Additionally, SNe~Ia that have statistical
evidence for circumstellar gas (as well as those with variable narrow
absorption features which provide strong evidence for circumstellar
gas) also have higher ejecta velocities on average
\citep{Foley12:csm}.  These results indicate that the progenitor
system or possibly orientation effects are related to the inferred
dust properties.

\subsection{SN~2014J}

SN~2014J was discovered at an $R$-band magnitude of 10.5 on 2014 Jan
21.805 (UT dates are used throughout this paper) by \citet{Fossey14}.
After its initial discovery, multiple groups reported pre-discovery
detections and limits \citep[e.g.,][]{Denisenko14, Dhungana14,
  Gerke14, Ma14, Zheng14}.  The SN was promptly spectroscopically
identified as a young SN~Ia \citep{Ayani14, Cao14, Itoh14}.  We
triggered multiple programs to study the photometric and spectroscopic
evolution of the SN, its circumstellar environment, its polarisation,
its energetics, and other aspects.  In particular, we triggered our
{\it Hubble Space Telescope} (\textit{HST}) target-of-opportunity
program to obtain ultraviolet (UV) spectra of SNe~Ia (GO-13286; PI
Foley; \citealt{Foley14:14jhst}).

SN~2014J, in M82, is the nearest SN~Ia in at least 28 years.  Our best
distance estimate of M82 (Section~\ref{ss:direct}) is $D = 3.3$~Mpc,
placing SN~2014J formally closer than SNe~1972E and 1986G ($D = 3.6$
and 3.7~Mpc), but uncertainties in the distance measurements currently
prevent a definitive ranking.  However, if these measurements are
accurate, SN~2014J is the nearest detected SN~Ia since Kepler's SN
(410~years).

Being so close, SN~2014J has been observed extensively at many
wavelengths.  \citet{Zheng14} presented early optical photometry of
SN~2014J, constraining the explosion time.  Nondetections in X-rays
\citep{Margutti14} and radio \citep{Chandler14, Chandra14, Chomiuk14}
provide no evidence for a dense and smooth circumstellar environment.
Pre-explosion optical images with no luminous source at the position
of SN~2014J are inconsistent with red supergiant companion stars
\citep{Goobar14, Kelly14}, while nondetections in pre-explosion X-ray
images are inconsistent with the progenitor system being in a
super-soft state in the decade before explosion \citep{Maksym14,
  Nielsen14}.

Spectropolarimetric observations of SN~2014J were made to directly
constrain $R_{V}$ by measuring the wavelength of maximum polarisation
\citep{Patat14}, indicating $R_{V} < 2$.  High-resolution spectroscopy
has revealed a multitude of interstellar and/or circumstellar
absorption features \citep{Welty14}, which have been used to detect
circumstellar material (CSM) in other SNe~Ia \citep{Patat07:06x,
  Blondin09, Simon09}.

\citet{Goobar14} presented optical and near-infrared (NIR) photometry,
several low-resolution optical spectra, and a high-resolution
spectrum.  Examining their light curves and spectra, they determine
that a simple reddening law with $R_{V} \approx 1.4$ provides a good
match to their data.  \citet{Marion14}, with similar data, came to
similar conclusions.  Recently, \citet{Amanullah14} added more
photometry, covering more bands and a longer time baseline to the
\citet{Goobar14} photometry, and also found a best-fitting value of
$R_{V} = 1.4 \pm 0.1$ for a simple dust reddening law.  They also
found that a power-law reddening law with an index of $-2.1 \pm 0.1$
is consistent with the data.

\begin{figure*}
\begin{center}
\includegraphics[angle=0,width=6.6in]{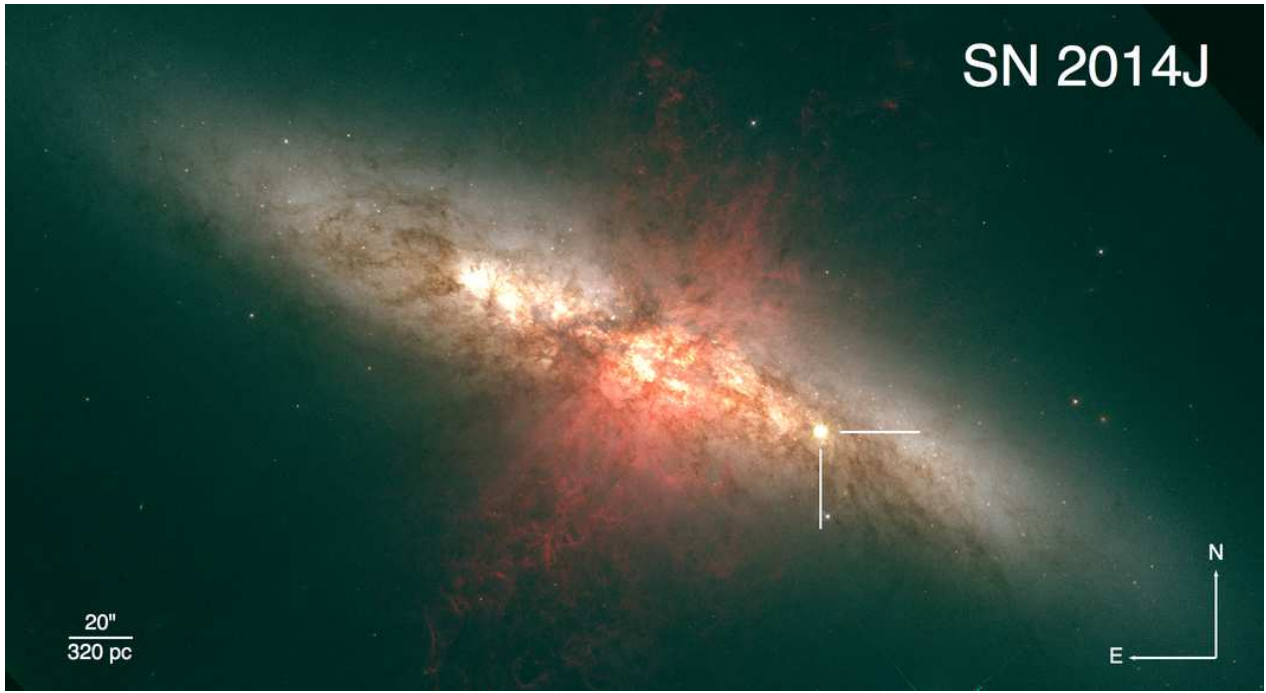}
\caption{{\it HST}/WFC3 image of SN~2014J in M82.  The RGB channels
  correspond to F658N+F814W, F555W, and F435W (roughly
  H$\alpha$+$IV\!B$), respectively.  Because the {\it HST} images of
  SN~2014J do not probe the wings of the PSF in the same way as the
  deep M82 pre-explosion image, simple stacking of the pre-explosion
  M82 and SN~2014J images produces an image where SN~2014J appears
  fainter than it should.  Using the brightness measurements of
  SN~2014J in these bands, an artificial star was generated with Tiny
  Tim to match the PSF of the deep, wide-field M82 image.  This source
  was inserted at the position of SN~2014J to create an accurate
  visualisation.}\label{f:finder}
\end{center}
\end{figure*}

In this manuscript, we present our UV, optical, and NIR data in
Section~\ref{s:obs}.  The extinction to SN~2014J is estimated in
Section~\ref{s:ext}.  We describe dust reddening and circumstellar
scattering models in Section~\ref{s:models}, and we use those models
to estimate the reddening of SN~2014J based on our photometry
(Section~\ref{s:phot_red}) and spectroscopy
(Section~\ref{s:spec_red}).  We discuss our findings and summarize our
conclusions in Section~\ref{s:disc}.


\section{Observations}\label{s:obs}

\subsection{Photometry}

SN 2014J in M82 was observed with {\it HST}/WFC3 UVIS over 7 epochs
between 2014 January 28 and 2014 March 07 (DD-13621; PI Goobar).  All
7 epochs include observations in the F218W, F225W, F275W, and F336W
filters.  Epochs 1 and 3 include observations in the F467M, F631N, and
F845M filters.  Epochs 2, 4, 5, and 6 include observations in F438W,
F555W, and F814W (roughly $B$, $V$, and $I$).

We combined exposures and performed cosmic-ray rejection using
AstroDrizzle after we performed the pixel-based charge-transfer
efficiency correction.  We registered the individual flatfielded (flt)
frames using TweakReg in DrizzlePac.  In the images, the SN was the
only detected object, so we did not attempt to register the absolute
astrometry or perform background subtraction.

An image combining {\it HST}/WFC3 observations of SN~2014J with deep
pre-explosion images of M82 is shown in Figure~\ref{f:finder}.  To
create this image, we obtained images of M82 from the Hubble Legacy
Archive observed in the F435W, F555W, F658N, and F814W filters
(roughly $B$, $V$, H$\alpha$, and $I$), with exposure times of 10,800,
8160, 26,400, and 4200~s, respectively.  Conversely, the SN images are
extremely short (0.48~s).  While this choice prevents saturated
images, the short exposure times also prevent an accurate
characterisation of the wings of the point-spread function (PSF).  As
such, a simple combination of the pre-explosion and SN images causes
the SN to appear much fainter to the human eye than its true
brightness.  Instead, we injected a model PSF, generated using Tiny
Tim \citep{Krist11}, at the location of the SN with the measured
brightness in each band.  We then combined the final images with
F435W, F555W, and F658N~+~F814W as the blue, green, and red channels.
SN~2014J still appears somewhat faint in Figure~\ref{f:finder} because
of the excellent {\it HST} PSF.  To see faint structures in M82, we
chose a dynamic range which saturates the SN.

We performed aperture photometry on the SN using the APPHOT package in
IRAF\footnote{IRAF: the Image Reduction and Analysis Facility is
  distributed by the National Optical Astronomy Observatory, which is
  operated by the Association of Universities for Research in
  Astronomy (AURA) under cooperative agreement with the National
  Science Foundation (NSF).}.  For each image, we used a 0\arcsec\!.4
aperture.  For the last epoch in F218W and the last two epochs of
F275W, the SN had faded enough such that using such a large aperture
was introducing a systematic bias into our results.  To account for
this, we measured the photometry for a 5-pixel (0\arcsec\!.2) aperture
and used aperture corrections derived from the earlier epochs to
convert to the 0\arcsec\!.4 aperture.  Our photometric uncertainties
include the reported uncertainty from Phot (assuming a read-noise
contribution of 3.15 electrons) and the uncertainty from the aperture
correction added in quadrature.  The uncertainty in the aperture
correction was taken to be the standard deviation of the measurements
from the early epochs.

We list all {\it HST} photometry in Table~\ref{t:hstphot}, but we only
include the F218W, F225W, F275W, and F336W data in our analysis.  The
other filters overlap with other bands which have more complete
temporal coverage.

The {\it HST} images were independently reduced and analysed by
\citet{Amanullah14}.  Comparing our measurements with those listed in
that work, we find the photometry to be consistent within the quoted
uncertainties.

Filtered CCD images of SN 2014J were obtained with the 0.76-m Katzman
Automatic Imaging Telescope \citep[KAIT;][]{Filippenko01} at Lick
Observatory All KAIT images were reduced using our image-reduction
pipeline \citep{Ganeshalingam10}.  PSF photometry was then performed
using DAOPHOT \citep{Stetson87}.  The SN instrumental magnitudes have
been calibrated to two nearby stars from the APASS
catalogue\footnote{http://www.aavso.org/apass/}.
The APASS magnitudes are transformed into the Landolt
system\footnote{http://www.sdss.org/dr7/algorithms/sdssUBVRITransform.html\\\#Lupton2005.}
before they are used for calibrating KAIT data.  Owing to the lack of
template images prior to the SN explosion, we have not performed image
subtraction.  Consequently, the photometry presented here should be
considered preliminary; however, because SN~2014J is much brighter
than its surrounding regions and the pre-explosion {\it HST} images do
not indicate a bright region coincident with SN~2014J, the background
contamination should be minimal for all epochs presented here.
Table~\ref{t:kait} lists the KAIT photometry.

NIR (\jhks\!\!) observations were made with FanCam, a 1024$\times$1024
pixel HAWAII-I HgCdTe imaging system \citep{Kanneganti09} on the
University of Virginia's 31-inch telescope at Fan Mountain, just
outside of Charlottesville, VA.  Observations consist of a series of
either 4, 8, or 10~s integrations.  We employed standard NIR
data-reduction techniques in IRAF.  The brightest parts of M82 can be
fit into a single array quadrant, so that dithering can efficiently
utilise empty quadrants as sky exposures.  Each quadrant was reduced
separately and, ultimately, coadded into a single image.  The data
were analysed with the online astrometry programs SEXTRACTOR and
SWARP.  Calibration was performed using field stars with reported
fluxes in the Two Micron All Sky Survey \citep[2MASS;][]{Skrutskie06}.
Table~\ref{t:fancam} lists the NIR photometry.

We report photometric parameters (times of maximum brightness, peak
brightness, decline rate) in Table~\ref{t:params}.  Importantly, we
measure $t_{\rm max}(B) = 2$,456,690.5 $\pm 0.2$ (JD), $V_{\rm max} =
10.61 \pm 0.05$~mag, and $\Delta m_{15} (B) = 0.95 \pm 0.01$~mag.  Our
computed values are consistent with those of \citet{Tsvetkov14} and
\citet{Marion14}.  We caution that these are the observed values and
are uncorrected for reddening.

\begin{figure}
\begin{center}
\includegraphics[angle=90,width=3.2in]{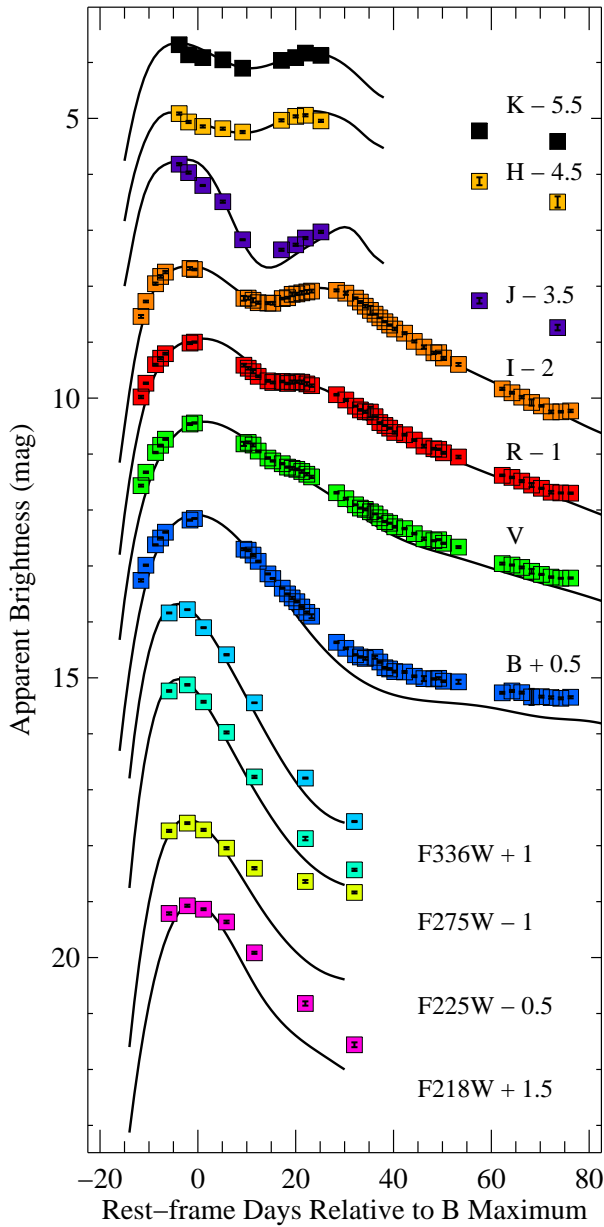}
\caption{{\it HST}/WFC3, KAIT, and FanCam ``UV,'' optical, and NIR
  light curves of SN~2014J (squares).  Overplotted are the optical and
  NIR light curves of SN~2011fe \citep[black curves;][]{Matheson12,
    Richmond12}, and the synthesised UV light curves from the
  SN~2011fe {\it HST} spectra, shifted to match the peak brightness of
  SN~2014J in each band.}\label{f:lc}
\end{center}
\end{figure}

All photometry is presented in Tables~\ref{t:hstphot} --
\ref{t:fancam} and the subset used for our analysis is shown in
Figure~\ref{f:lc}.

\subsection{Spectroscopy}

SN~2014J was observed by \textit{HST} using the STIS spectrograph on
ten epochs from 2014 January 26.60 to 2014 February 26.07,
corresponding to $t= -6.4$ to 24.1~days relative to $B$-band maximum.
Each individual spectrum was obtained over two to five orbits with the
$52\arcsec \times 0.\arcsec2$ slit and three different setups: the
near-UV MAMA detector and the G230L grating, the CCD/G430L, and the
CCD/G750L.  The three setups yield a combined wavelength range of
1615--10,230~\AA.  A log of observations is presented in
Table~\ref{t:hstspec}.

\begin{figure}
\begin{center}
\includegraphics[angle=0,width=3.2in]{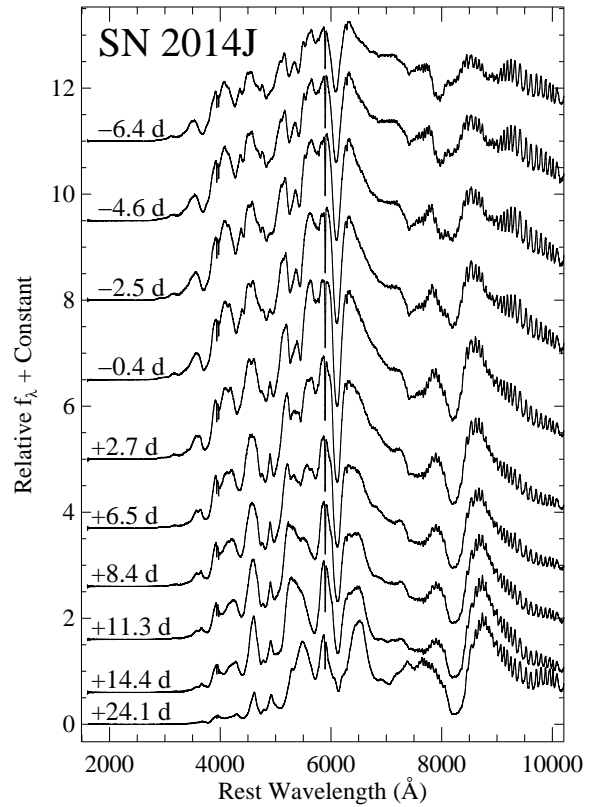}
\caption{{\it HST}/STIS spectral time series of SN~2014J.  The phase
  relative to $B$-band maximum brightness is labeled.  The spectra
  have been dereddened only by the Milky Way reddening
  estimate.}\label{f:hst_all}
\end{center}
\end{figure}

The data were reduced using the standard \textit{HST} Space Telescope
Science Data Analysis System (STSDAS) routines to bias subtract,
flatfield, extract, wavelength-calibrate, and flux-calibrate each SN
spectrum.  The spectra are presented in Figure~\ref{f:hst_all}.

Ground-based NIR spectroscopy of SN~2014J was obtained by the NASA
Infrared Telescope Facility (IRTF) with SpeX \citep{Rayner03} in
cross-dispersed mode and a 0\farcs3 slit, which yields 0.8--2.5 $\mu$m
wavelength coverage at $R = 2000$.  All observations were taken
utilising the standard ABBA technique, with the slit oriented along
the parallactic angle.  A telluric standard with a A0V spectral type
was observed for flux calibration and telluric correction.  The data
reductions made use of the Spextool and {\sc XTELLCOR} software
packages \citep{Vacca03, Cushing04}.  A log of observations is
presented in Table~\ref{t:nirspec}.  These data are also presented by
\citet{Marion14}.

High-dispersion optical spectroscopy of SN~2014J was obtained with the
Tillinghast Reflection Echelle Spectrograph \citep[TRES;][]{Furesz08}
on the Tillinghast 1.5~m telescope at the Fred Lawrence Whipple
Observatory.  TRES is a fiber-fed, crossed-dispersed spectrograph
covering the wavelength range 3850--9100~\AA\ with some gaps at
wavelengths beyond 6650~\AA.  Ten observations of SN~2014J were made
between 23 January 2014 and 22 March 2014 employing a $100~\mu$m
(2.3~\arcsec) fiber giving $R = 30$,000.  A log of observations is
presented in Table~\ref{t:tres}.


\section{Extinction Estimates}\label{s:ext}

Extinction and reddening measurements are well defined for stars and
galaxies, but SNe, with their broad spectral features, require more
attention to detail.  Here we define some necessary terms for
comprehending our results.

For an extinction ($A_{X}$), colour excess ($E(X-Y)$), or the ratio of
the total-to-selective extinction ($R_{\lambda}$), there are observed
and ``true'' quantities.  The observed quantities are defined as
\begin{align}
  A_{X}^{\rm obs}  &\equiv X^{\rm obs} - X^{\rm no~red}, \\
  E(X-Y)^{\rm obs} &\equiv A_{X}^{\rm obs} - A_{Y}^{\rm obs},~{\rm and} \\
  R_{V}^{\rm obs}  &\equiv A_{V}^{\rm obs} / E(B-V)^{\rm obs},
\end{align}
where $X^{\rm no~red}$ and $X^{\rm obs}$ are the brightness of the SN
in the $X$ band without any extinction and as observed, respectively.

Meanwhile, $E(X-Y)^{\rm true}$ and $R_{V}^{\rm true}$ are the
reddening and the ratio of total-to-selective extinction,
respectively, required such that dereddening the observed spectrum of
the SN results in a spectrum equivalent to the unreddened spectrum.

One can convert from the observed and true quantities by knowing the
underlying SED of the SN \citep[e.g.,][]{Phillips99, Nugent02}.
Because of a shift in the effective wavelength of photometric filters
with increased reddening (Section~\ref{ss:filt}), spectroscopy is
preferred for disentangling the differences between observed and true
parameters.

\subsection{Milky Way Reddening}

The nominal Milky Way reddening toward M82, as determined by
\citet{Schlafly11} from \citet{Schlegel98} dust maps, is $E(B-V) =
0.138$~mag.  This value is used by \citet{Goobar14} and several other
studies of SN~2014J.  \citet{Dalcanton09} note that M82 is a strong
source in the dust maps, biasing the reddening measurement.  They
instead suggest a value of $E(B-V) = 0.061$~mag as determined from
regions surrounding M82.  This value is used by \citet{Amanullah14}.
However, the \citet{Dalcanton09} estimate is on the \citet{Schlegel98}
scale; converting to the \citet{Schlafly11} scale, the Milky Way
reddening is $E(B-V) = 0.054$~mag, which we use in this work.

\subsection{Direct Measurement}\label{ss:direct}

SNe~Ia are standardisable candles in the optical and nearly standard
candles in the NIR.  One can directly measure the extinction to
SN~2014J from measuring its distance and brightness.

The distance to M82 is measured with moderate precision.  Parallel
observations of our STIS UV campaign were conducted to find and
measure distances from Cepheids in M82.  These data will be presented
elsewhere, but should provide a precise distance estimate for the
galaxy.

NED lists 7 distance estimates (with uncertainties) from 3 methods,
ranging from $\mu = 27.53 \pm 0.05$~mag \citep{Dalcanton09} to $\mu =
28.57 \pm 0.80$~mag \citep{Tully88}.  There are large discrepancies
even from the same methods and authors.  For instance,
\citet{Dalcanton09} measured tip of the red giant branch (TRGB)
distances of $\mu = 27.53 \pm 0.05$ and $27.74 \pm 0.04$ for two
positions in the galaxy.  These measurements are discrepant at the
3.3$\sigma$ level and correspond to a distance difference of 320~kpc.
M82 is in the M81 group and is interacting with M81 \citep{Yun94},
which has a Cepheid distance of $\mu = 27.60 \pm 0.03$~mag
\citep{Gerke11}.

For the NIR bands, SNe~Ia are nearly standard candles
\citep{Krisciunas04:hubble, Wood-Vasey08, Mandel09, Folatelli10,
  Burns11, Kattner12}.  SN~2014J appears to have a normal decline
rate, further indicating that it will have a standard peak luminosity
at these wavelengths.  Using the $K$-band light curve of SN~2014J, the
suite of peak $K$-band absolute magnitudes listed by
\citet{Matheson12}, and {\it no} extinction in M82, we find a weighted
average distance modulus of $\mu = 27.57 \pm 0.13$~mag and a limit of
$\mu < 27.60 \pm 0.08$~mag\footnote{We have no data before maximum in
  the NIR bands, but do cover the time of $B$-band maximum brightness.
  Some methods measure the absolute magnitude at peak in a given band,
  for which we only have limits, while others measure the absolute
  magnitude at the time of $B$-band maximum brightness.}.  This value
is only consistent with the closest distances measured for M82.  For
reasonable assumptions about the reddening and extinction law for
SN~2014J, we expect the extinction in the $K$ band, $A_{K}$, to be
$\lesssim 0.05$--0.15~mag, but any extinction in the $K$ band will
further decrease the distance modulus.  We therefore assume $\mu =
27.60$~mag, a value consistent with the $K$-band SN~2014J
measurements, the M81 Cepheid distance, and the closer direct distance
measurements of M82.  We adopt an uncertainty of 0.10~mag, which is
reasonable for the set of distances that are allowed by the $K$-band
data.

We measure the peak brightness of SN~2014J in all bands, and these are
reported in Table~\ref{t:params}.  Using the peak absolute magnitudes
from \citet{Prieto06} and the suite of measurements collated by
\citet{Matheson12} for the optical and NIR, respectively, we measure
the extinction in each band, also listed in Table~\ref{t:params}.
Assuming that SN~2014J had a typical SN~Ia luminosity, we find the
observed visual extinction to be $A_{V}^{\rm obs} = 2.07 \pm 0.18$~mag
at maximum brightness.

\subsection{Extinction from Optical-Infrared Colours}

The extinction in band $X$ ($A_{X}$) is essentially equivalent to
$E(X-Y)$ for cases where $A_{Y}^{\rm obs} \approx A_{Y}^{\rm true}
\approx 0$~mag.  If $A_{Y}$ is negligible, then this measurement is
independent of distance.  As shown above, $A_{K} \approx 0$~mag for
reasonable assumptions for the distance, and therefore $A_{V}^{\rm
  obs} \approx E(V-K)^{\rm obs}$.  Using a \citet{Cardelli89}
reddening law with $R_{V} = 3.1$, $A_{V}^{\rm obs} = 1.14 E(V-K)^{\rm
  obs}$.  We provide detailed measurements of $E(V-K)$ in
Section~\ref{ss:colex}.  Using the average reddening, we estimate
$A_{V}^{\rm obs} = 1.91 \pm 0.10$~mag.

We similarly determined that $A_{H}^{\rm obs}$ should be low
(0.13~mag), and we measure an average reddening of $E(V-H)^{\rm obs} =
1.77 \pm 0.12$~mag.  Therefore, $A_{V}^{\rm obs} = E(V-H)^{\rm obs} +
A_{H}^{\rm obs} = 1.91$~mag is consistent with our earlier
measurements.

From multiple methods, we determine that there is $A_{V}^{\rm obs} =
1.95 \pm 0.09$~mag of host-galaxy extinction along the line of sight
to SN~2014J.  Because of our methodology, this measurement is only
valid at maximum brightness.  Using the same methods, we measure
$A_{B}^{\rm obs} = 3.14 \pm 0.11$~mag.  Therefore, from
model-independent methods, we determine a maximum-brightness reddening
of $E(B-V)^{\rm obs} \equiv A_{B}^{\rm obs} - A_{V}^{\rm obs} = 1.19
\pm 0.14$~mag and $R_{V}^{\rm obs} \equiv A_{V}^{\rm obs} /
E(B-V)^{\rm obs} = 1.64 \pm 0.16$.

\subsection{Extinction from High-resolution Spectroscopy}

As discussed by \citet{Welty14}, the spectrum of SN~2014J displayed
strong, complex interstellar absorption lines of \ion{Na}{I},
\ion{K}{I}, \ion{Ca}{I}, \ion{Ca}{II}, CH, CH+, and CN, as well as a
number of diffuse interstellar bands (DIBs).  \citeauthor{Welty14}
interpreted the strong absorption at velocities $v_{\rm LSR} >
30$~\kms\ as arising in the interstellar medium (ISM) of M82.  We have
used the TRES spectra of SN~2014J, which cover the phases from $-9.6$
to $+45.3$~days with respect to the epoch of $B$-band maximum
brightness, to look for possible variability of the DIBs at 5870~\AA\
and 5797~\AA, and the \ion{Na}{I}~D and \ion{K}{I}~$\lambda$7665
absorption complexes.

\begin{figure}
\begin{center}
\includegraphics[angle=0,width=3.2in]{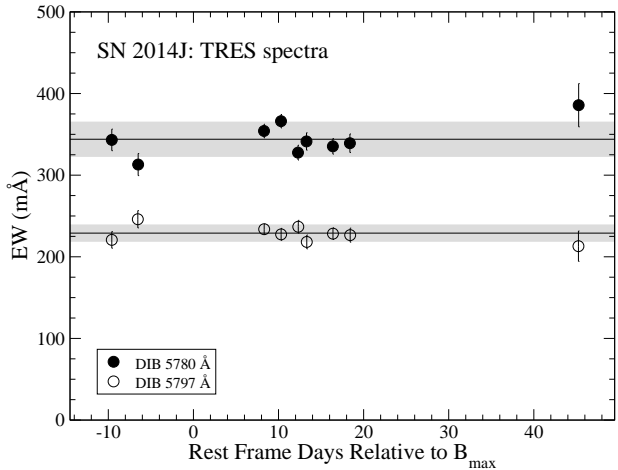}
\caption{Equivalent-width measurements of the DIB features at
  5780~\AA\ and 5797~\AA\ in the spectra of SN~2014J plotted as a
  function of light-curve phase.  The horizontal lines and grey
  regions represent the mean and 1$\sigma$ scatter for each feature.
  As noted in the text, these measurements are for the combined Milky
  Way~$+$~M82 absorption.  The final points at $+45$~days were
  measured from the sum of the spectra obtained on 2014 March 16 and
  2014 March 22.}\label{f:dib}
\end{center}
\end{figure}

The equivalent widths of the DIBs were calculated using the IRAF task
{\tt fitprofs} assuming Gaussian profiles of 2.8~\AA\ full width at
half-maximum intensity (FWHM) and 2.0~\AA\ FWHM, respectively.  The
measurements are displayed in Figure~\ref{f:dib}.  No variability is
detected for either DIB over the period of the observations.  Using
all data, we find weighted mean equivalent widths of EW(5780)~$= 344
\pm 21$~m\AA\ and EW(5797)~$= 229 \pm 10$~m\AA.  These numbers are for
the combined Milky Way~$+$~M82 absorption since it is not possible to
separate them.  However, as discussed by \citeauthor{Welty14}, the
Milky Way contribution is expected to be small (\about 5\%).  Our
measurements are in excellent agreement with those of
\citeauthor{Welty14}, but the EW(5780) value quoted by
\citet{Goobar14} of $480 \pm 10$~~m\AA\ from high-dispersion spectra
obtained on 26~January~2014 and 28~January~2014 (UT) ($-6$ and
$-4$~days with respect to $B$~maximum) differs by 5.8$\sigma$.  Using
our measurement of EW(5780), a total visual extinction of $A_{V} = 1.8
\pm 0.9$~mag is inferred from Equation~6 of \citet{Phillips13}.

\begin{figure}
\begin{center}
\includegraphics[angle=0,width=3.2in]{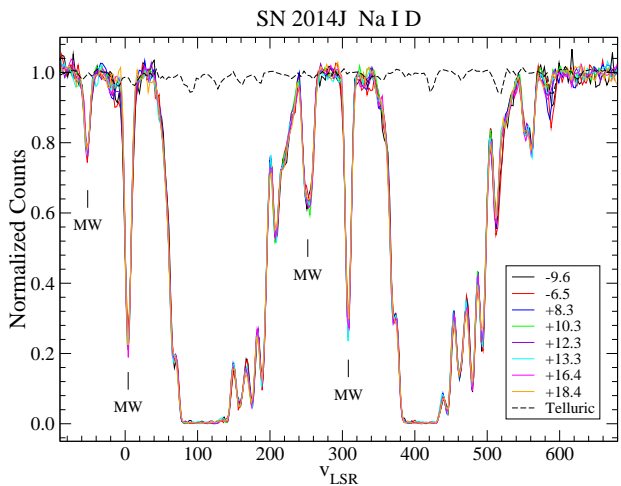}
\caption{Normalised profiles of the \ion{Na}{I}~D line absorption in
  SN~2014J from $-9.6$ to $+18.4$~days with respect to the epoch of
  $B$-band maximum brightness.  The different epochs have different
  colours and are labeled.  The telluric spectrum, as measured from a
  hot star, is plotted as a dashed line.  The absorption components
  likely associated with Milky Way gas are marked.  The remaining
  absorption is likely from the ISM of M82 or the CSM of
  SN~2014J.}\label{f:na}
\end{center}
\end{figure}

Figure~\ref{f:na} shows the evolution of the \ion{Na}{I}~D absorption
in SN~2014J from $-9.6$ to $+18.4$~days with respect to $B$-band
maximum brightness. (The final two TRES spectra obtained 1.5 months
after $B$~maximum are not included in this figure owing to their lower
S/N.)  The telluric absorption features at these wavelengths are also
indicated.  No credible evidence is seen in the \ion{Na}{I}~D lines
for variations of any of the absorption components, although the
strongly saturated absorption at $+60 > v_{\rm LSR} >
+150$~km~s$^{-1}$ would make small variations in this velocity range
difficult to discern.  The weaker \ion{K}{I}~$\lambda\lambda$7665,
7699 lines are much more useful for exploring these absorption
components.  Our TRES spectra, which covered the 7665~\AA\ line only,
also show no evidence for variations.

\section{Reddening/Scattering Models}\label{s:models}

To further understand the reddening and scattering properties of
SN~2014J, we apply specific models to our photometric and
spectroscopic data.  We have chosen 6 models to test, as follows.

\begin{itemize}
\item {\bf CCM31} A \citet[hereafter,
  \citetalias{Cardelli89}]{Cardelli89} reddening law as modified by
  \citet{Odonnell94} with $R_{V} = 3.1$, the canonical Milky Way
  value.

\item {\bf F9931}: A \citet[hereafter,
  \citetalias{Fitzpatrick99}]{Fitzpatrick99} reddening law with $R_{V}
  = 3.1$.

\item {\bf CCM}: A \citet{Cardelli89} reddening law as modified by
  \citet{Odonnell94} with no restriction on $R_{V}$.

\item {\bf F99}: A \citet{Fitzpatrick99} reddening law with no
  restriction on $R_{V}$.  The values reported here use the Milky Way
  dust parameters for this reddening law.  We also fit the data with
  an LMC reddening law, which typically changed any parameter by
  $<$2\%.

\item {\bf CSM}: A circumstellar material scattering model as
  described by \citetalias{Goobar08}.  Here the reddening is described
  by a power law.

\item {\bf CSMD}: A two-component model with both a circumstellar
  scattering component and a \citetalias{Fitzpatrick99} dust reddening
  component.

\end{itemize}

The scattering of SN photons off circumstellar dust should alter the
observed light curves and spectra from SNe~Ia \citep{Wang05, Patat06,
  Goobar08, Amanullah11}.  \citetalias{Goobar08} simulated this effect
and determined the effects on the observed extinction,
\begin{equation}
  \frac{A_{\lambda}}{A_{V}} = 1 - a + a \left ( \frac{\lambda}{\lambda_{V}} \right )^{p}, \label{e:g08}
\end{equation}
where $\lambda_{V}$ is the wavelength of the $V$ band (chosen to be
5500~\AA\ by \citetalias{Goobar08}), and $a$ and $p$ are free
parameters.  This relation can be rewritten as an effective $R_{V}$,
\begin{equation}
  R_{V} = \frac{1}{a (0.8^{p} - 1)}.
\end{equation}
For scattering from LMC (Milky Way) dust, \citetalias{Goobar08}
measured $a = 0.9$ ($a = 0.8$) and $p = -1.5$ ($p = -2.5$),
respectively, corresponding to $R_{V} = 1.67$ (2.79).

Circumstellar scattering produces a light echo that is delayed by the
light travel time between the SN and the CSM \citep{Wang05, Patat06,
  Amanullah11}.  For photons arriving at the same time, the light echo
and SN SEDs will be different, resulting in temporal differences in
spectral features and a changing continuum \citep{Patat06,
  Amanullah11}.  The details of how the time-delayed SED contributes
to the observed data are not included in our scattering
parameterisation (Equation~\ref{e:g08}).


\section{Photometric Reddening Estimates}\label{s:phot_red}

Spectroscopy provides precise measurements of wavelength-dependent
reddening and extinction.  Photometry can provide high temporal
resolution and potentially probe wavelengths difficult to measure with
spectra.

\subsection{Comparison Supernovae}\label{ss:comp}

We presented the SN~2014J light curves in Section~\ref{s:obs}, and
Figure~\ref{f:lc} displays those data.  In that figure, we compare the
SN~2014J light curves to those of SN~2011fe.  Below, we also compare
to SNe~2009ig and 2013dy.  With the exception of SN~2011iv
\citep{Foley12:11iv} and SN~2014J, SNe~2011fe and 2013dy are the only
SNe~Ia with high-quality UV spectral time series from {\it HST}.
\citet{Maguire12} presented the optical {\it HST} spectra of
SN~2011fe, \citet{Foley13:ca} presented the maximum-light UV-optical
spectrum, and \citet{Mazzali14} presented the full spectral series.
The SN~2013dy spectra will be studied in detail in another publication
(Pan et~al., in preparation).  UV spectra of SN~2009ig were obtained
by {\it Swift} \citep{Foley12:09ig}.  Although these data are not as
high quality as the {\it HST} data, the small number of well-observed
SNe~Ia with {\it HST} necessitate their inclusion in this study.

We used SNID \citep{Blondin07} to determine which SNe were spectrally
similar to SN~2014J.  SNID removes the continuum from each spectrum,
and thus reddening effects are reduced.  The overwhelming match for
SN~2014J was SN~2007co; spectra of SN~2007co were in the top 5 SNID
matches for our 10 {\it HST} spectra 16 times (32\%).  Although this
is a crude metric that depends on the number of spectra for a given SN
at specific epochs in the SNID database, it does indicate that
SN~2007co is spectrally similar.  Of our comparison SNe with UV
spectra, SN~2007co is spectrally most similar to SN~2011fe, and their
colour curves have similar evolution, although SN~2007co is about
0.07~mag redder in $B-V$ than SN~2011fe at all epochs.

SN~Ia intrinsic colours correlate with both light-curve shape
\citep[e.g.,][]{Riess96} and velocity \citep{Foley11:vel, Foley12:vel,
  Mandel14}.  It is likely that even for two SNe with exactly the same
light-curve shape and velocity, some intrinsic colour scatter exists.
In fact, this has been shown for SNe~2011by and 2011fe
\citep{Foley13:met}.  None the less, by comparing SN~2014J to several
similar SNe, we hope to probe most of the possible parameter space.

SN~2014J is a somewhat slow decliner ($\Delta m_{15} (B)_{\rm obs} =
0.95$~mag), similar to SNe~2009ig ($\Delta m_{15} (B)_{\rm obs} =
0.89$~mag; \citealt{Foley12:09ig}) and 2013dy ($\Delta m_{15} (B)_{\rm
  obs} = 0.86$~mag; W.\ Zheng, 2014, private communication).
SNe~2007co and 2011fe are slightly faster decliners, with $\Delta
m_{15} (B)_{\rm obs} = 1.16$ and 1.10~mag, respectively
\citep[e.g.,][]{Mandel11, Richmond12}.  However, the measurement of
$\Delta m_{15} (B)$ is affected by dust reddening (as the effective
wavelength of the filter shifts).  \citet{Phillips99} presented a
method to derive a corrected value:
\begin{equation}
  \Delta m_{15} (B)_{\rm true} = \Delta m_{15} (B)_{\rm obs} + 0.1 E(B-V).\label{e:dm15}
\end{equation}
Below, we argue that $E(B-V) \approx 0.6$--1.3~mag for SN~2014J, which
would result in $\Delta m_{15} (B)_{\rm true} = 1.01$--1.08~mag.
Similarly, SN~2013dy had moderate reddening, both from its host galaxy
and the Milky Way, resulting in $\Delta m_{15} (B)_{\rm true} =
0.89$~mag.  SNe~2007co, 2009ig, and 2011fe do not suffer from
significant extinction \citep{Mandel11, Nugent11, Foley12:09ig,
  Johansson13, Patat13}, and the observed decline rate is essentially
equivalent to the true decline rate.  Therefore, SN~2014J has a
decline rate intermediate to SNe~2009ig/2013dy and SNe~2007co/2011fe,
but all are similar.

SN~2014J has a relatively high ejecta velocity at maximum brightness.
From the $t = -0.4$~day spectrum of SN~2014J, we measure a
\ion{Si}{II} $\lambda 6355$ velocity of $-11$,870~\kms.  From the $t =
-0.4$~day spectrum of SN~2013dy, we measure a \ion{Si}{II} $\lambda
6355$ velocity of $-10$,370~\kms.  SNe~2007co, 2009ig, and 2011fe have
maximum-light \ion{Si}{II} $\lambda 6355$ velocities of $-12$,000
\citep{Foley11:vel}, $-13$,500 \citep{Foley12:csm}, and
$-10$,400~\kms\ \citep{Foley13:met}, respectively.  Therefore,
SN~2014J has ejecta velocity intermediate to that of SNe~2011fe/2013dy
and SN~2009ig, and similar to SN~2007co.  Since the intrinsic $B-V$
colours (and presumably flux at shorter wavelengths) of SNe~Ia are
correlated with ejecta velocity \citep{Foley11:vel, Foley11:vgrad,
  Mandel14}, spanning the range of possible velocities is important.

Although SN~2014J has a longer rise and decline in bluer bands, the
redder bands, especially $I$ and $H$, are nearly identical for
SNe~2011fe and 2014J (Figure~\ref{f:lc}).  Therefore, a single stretch
cannot describe the light-curve differences.  Instead, the observed
differences in stretch are likely caused by different effective
wavelengths (which change with time).  Both scattering and dust
extinction will affect the bluer bands more, leading to a larger
discrepancy in those bands.

Before understanding the reddening for SN~2014J, we must understand
(and correct for) the reddening for our comparison SNe.  First, we
remove all Milky Way reddening as determined by \citet{Schlafly11}
using a \citet{Fitzpatrick99} reddening law and $R_{V} = 3.1$.  For
SNe~2007co, 2009ig, and 2011fe, the host-galaxy reddening is estimated
to be minimal \citep[e.g.,][]{Mandel11, Patat11, Foley12:09ig,
  Johansson13, Phillips13}\footnote{Although SN~2009ig appears to have
  no dust reddening, a light echo has been detected for the SN,
  indicating that there could be substantial dust along the line of
  sight to SN~2009ig \citep{Garnavich13}.}.  We make no host-galaxy
dust correction for these SNe.

SN~2013dy likely suffers from some host-galaxy dust reddening.
\citet{Zheng13} estimate a reddening of $E(B-V) = 0.15$~mag based on
measurements of the Na~D lines.  However, this method is not
particularly accurate \citep{Blondin09, Poznanski11, Phillips13}.  To
account for the possible uncertainty, we deredden SN~2013dy assuming
$E(B-V) = 0.15 \pm 0.10$~mag.  For SN~2013dy, we use all combinations
of reddening, $R_{V}$ from 1.0 to 3.1, and both
\citetalias{Fitzpatrick99} and \citetalias{Cardelli89} reddening laws;
the differences between approaches are included in the overall
uncertainty for any results.

\subsection{Photometric Filters}\label{ss:filt}

Knowing the effective wavelength of photometric filters is especially
important when inferring reddening properties from photometry.  In the
UV, the intrinsic flux of a SN~Ia decreases to shorter wavelengths,
and even a modest amount of reddening can dramatically alter the
effective wavelength of some filters.  For the WFC3
filters\footnote{ftp://ftp.stsci.edu/cdbs/comp/wfc3/}, this is
compounded by ``red leaks,'' where the transmission of the filters
extends (at a low level) far into the optical and NIR.

\begin{figure}
\begin{center}
\includegraphics[angle=90,width=3.2in]{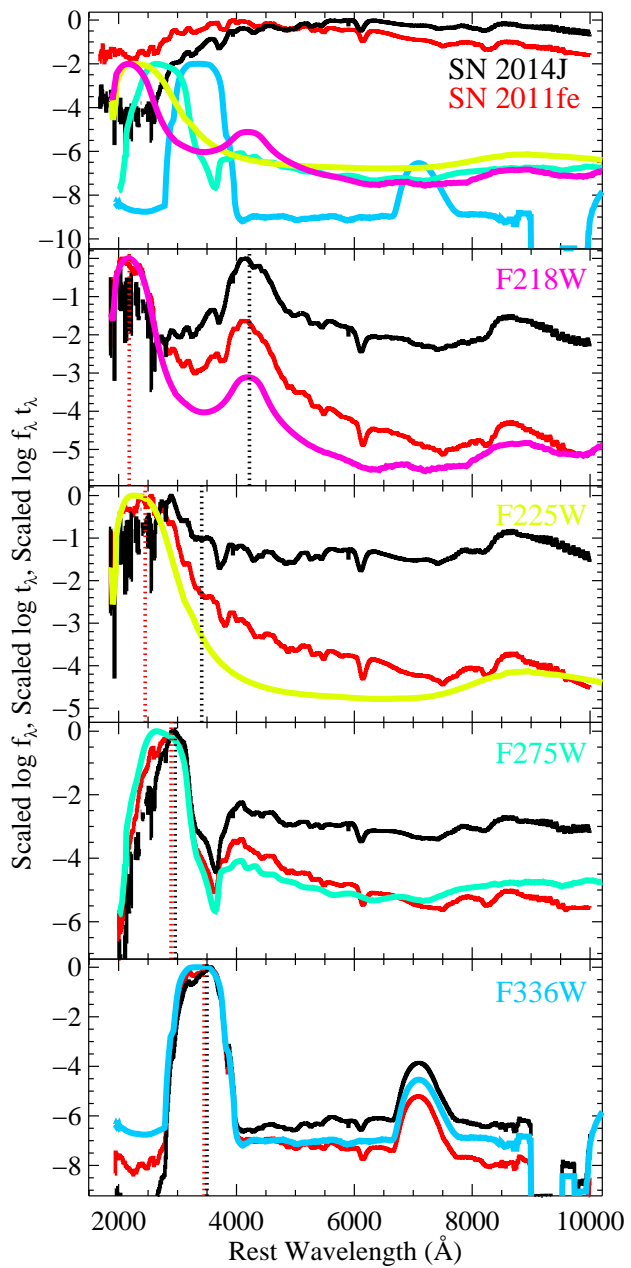}
\caption{({\it Top panel}): Maximum-brightness spectra of SNe~2014J
  (black) and 2011fe (red) scaled such that their peak flux is 1
  ($\log f_{\lambda} = 0$).  The magenta, yellow, mint, and light blue
  curves are the WFC3 F218W, F225W, F275W, and F336W transmission
  curves, respectively, also scaled such that their peak is 0.01
  ($\log t_{\lambda} = -2$).  ({\it Bottom panels}): From top to
  bottom, the F218W, F225W, F275W, and F336W transmission curves
  scaled such that their peak is 1 ($\log t_{\lambda} = 0$) in the
  colours labeled above.  The solid red and black curves are the
  transmitted flux (the flux times the transmission curve) for
  SNe~2011fe and 2014J, respectively, scaled such that their peak is 1
  ($\log f_{\lambda} t_{\lambda} = 0$).  The dotted red and black
  lines represent the effective wavelength in each filter for
  SNe~2011fe and 2014J, respectively.}\label{f:spec_filt}
\end{center}
\end{figure}

Figure~\ref{f:spec_filt} shows the observed maximum-brightness spectra
of SNe~2011fe and 2014J, the transmission curves for the WFC3 filters,
and the transmitted spectra of the SNe.  We measure the effective
wavelength of each filter for both SNe.  While the effective
wavelengths for the two SNe in F275W and F336W are relatively similar
(differing by 70 and 50~\AA, respectively), the F218W and F225W
filters have significantly different effective wavelengths with
differences of 2040 and 960~\AA, respectively.  As a result of these
shifts, these ``bluer'' filters have an effectively longer wavelength
than F275W.

None of the {\it HST} photometry exclusively probes UV wavelengths for
SN~2014J. The effectively bluest filter, F275W, has an effective
wavelength of 2960~\AA\ at maximum brightness, but a significant
fraction of the photons come from the optical.  Since SNe~Ia become
redder after maximum brightness, the effective wavelengths of the
``UV'' filters all shift farther to the red.  Although these are
nominally UV filters, the majority of SN~2014J photons measured
through these filters had optical wavelengths.  Only our {\it HST}
spectroscopy effectively probes the UV properties of SN~2014J.

There is a significant systematic uncertainty related to the effective
wavelength for these filters.  More than 50\% of the flux for the
F218W and F225W filters occurs where the transmission function is
$<$0.1\% of the peak transmission.  Slight uncertainties in the filter
transmission curves at these wavelengths will cause large
uncertainties in effective wavelengths {\it and} any synthetic
photometry.

To test the potential systematic uncertainty, we synthesised
photometry from our spectra of SN~2014J and compared the results to
the photometry.  For our optical filters, the synthesised photometry
was consistent with the observed photometry to within the photometric
uncertainties.  However, the UV synthesised photometry was brighter
than the observed photometry by roughly 0.15, 0.52, 0.18, and 0.10~mag
(the exact amount changes as the spectrum evolves) from the bluest to
the reddest bands, respectively.  Although our spectra are somewhat
noisy at the shortest wavelengths, Figure~\ref{f:spec_filt} shows that
the majority of the flux in these filters is coming from $\lambda
\gtrsim 3000$~\AA, where our spectra have a high S/N.  These
differences are most likely caused by inaccurate filter transmission
curves.

The difference in effective wavelength even leads to a systematic
uncertainty related to the Milky Way extinction.  The difference in
the extinction for the nominal and effective wavelength of the F218W
filter, for instance, is $>$0.1~mag for SN~2014J, which has a
relatively small Milky Way reddening of $E(B-V) = 0.05$~mag.  The
exact value for the extinction is extremely uncertain even if one has
perfect knowledge of the SED simply because of the uncertainty in the
filter transmission.

Because of these systematic effects, the F218W and F225W bands should
{\it never} be used for measuring dust properties of even moderately
reddened SNe~Ia.  The only way to properly probe these UV wavelengths
with current instrumentation is with {\it HST} spectroscopy.  For
determining the dust and scattering properties for SN~2014J, which is
quite reddened, we still use the F275W and F336W photometry, but
include systematic uncertainties of 0.18 and 0.10~mag, respectively.

\subsection{Colour Excess}\label{ss:colex}

Since SN~2011fe had essentially no host-galaxy reddening
\citep{Nugent11, Johansson13, Patat13}, we can approximate $(X-Y)_{\rm
  11fe}^{\rm obs} = (X-Y)_{\rm 11fe}^{\rm true}$.  One can then
determine the observed colour excess to SN~2014J by comparing colour
curves:
\begin{equation}
  E(X-Y)_{\rm 14J}^{\rm obs} = (X-Y)_{\rm 14J}^{\rm obs} - (X-Y)_{\rm 11fe},
\end{equation}
where $X$ and $Y$ are any two given photometric bands.  This method
assumes similar SEDs and spectral evolution.

We use the \citet{Richmond12} and \citet{Matheson12} optical and NIR
SN~2011fe light curves.  We also synthesise UV light curves from the
SN~2011fe UV spectral sequence \citep{Mazzali14}.  For convenience, we
provide these synthetic light curves in Table~\ref{t:11fe_uv}.

\begin{figure*}
\begin{center}
\includegraphics[angle=90,width=6.4in]{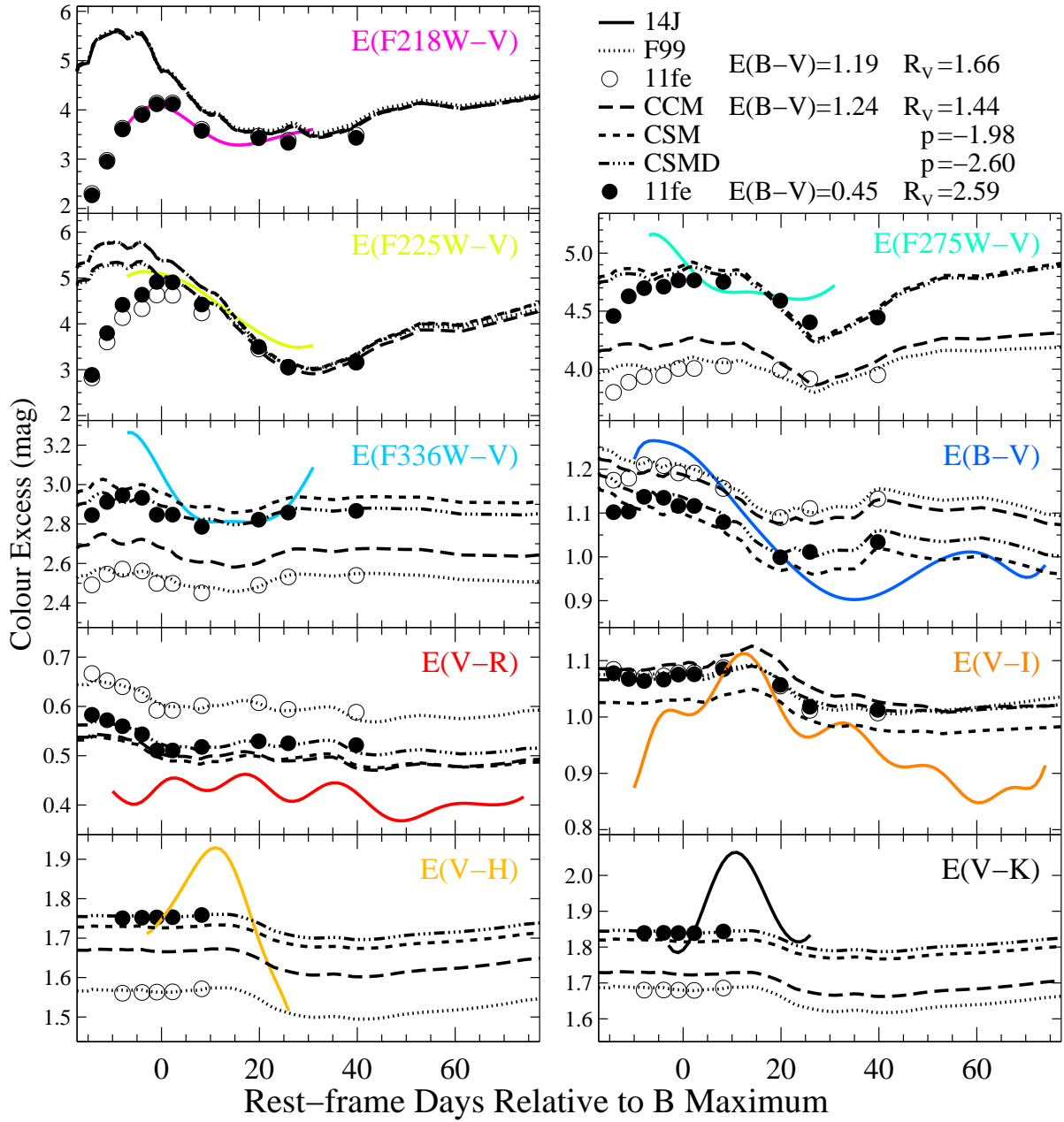}
\caption{Observed colour excesses for SN~2014J (solid curves).  Each
  colour excess is determined from the photometry of SN~2014J relative
  to that of SN~2011fe, and the plotted colour corresponds to the
  colour of the non-$V$ band filter used for the colour excess as
  displayed in Figure~\ref{f:lc}.  Since reddening and scattering
  change the effective wavelength of each filter and SN SEDs change
  dramatically with time, the measured colour excess is expected to
  change in some filters with time.  To account for this change, we
  also present spectrophotometry of template spectra \citep{Hsiao07}
  reddened and/or scattered by various amounts.  Four scenarios are
  shown: \citetalias{Fitzpatrick99} reddening with $E(B-V) = 1.19$~mag
  and $R_{V} = 1.66$ (dotted curve), \citetalias{Cardelli89} reddening
  with $E(B-V) = 1.24$ and $R_{V} = 1.44$ (long-dashed curve), CSM
  scattering with $p = -1.98$ (short-dashed curve), and a combined
  \citetalias{Fitzpatrick99} reddening with $E(B-V) = 0.45$~mag and
  $R_{V} = 2.57$ and CSM scattering with $p = -2.57$ (CSMD; dot-dashed
  curve).  The same procedure was done for spectra of SN~2011fe (open
  circles corresponding to the \citetalias{Fitzpatrick99} reddening
  and filled circles corresponding to the CSMD model).}\label{f:colex}
\end{center}
\end{figure*}

We interpolate light curves using B-splines.  This method, rather than
fitting a simple polynomial or a template light curve, is relatively
assumption free and provides an excellent description of the data with
small residuals.  From the interpolated light curves, we derived
colour curves and then colour-excess curves.  We present the
$E(X-V)^{\rm obs}$ and $E(V-Y)^{\rm obs}$ colour-excess curves in
Figure~\ref{f:colex}.

Examining Figure~\ref{f:lc}, the $J$-band light curves for SNe~2011fe
and 2014J have different shapes.  Since this behaviour is not seen in
the $I$ or $H$ bands, we believe this is related to the filter
transmission functions for the $J$ band \citep[$S$
corrections;][]{Stritzinger02}.  \citet{Matheson12} note that their
$J$-band light curve of SN~2011fe deviates significantly from template
light curves, likely because of the significantly different $J$ filter
used for that study.  Accordingly, we do not include any $J$-band data
in this analysis.

The observed colour excesses for SN~2014J change dramatically with
time for some colours.  For instance $E(B-V)$, the preferred colour
excess to describe the amount of reddening, changes by \about 0.4~mag
from maximum brightness to one month after maximum brightness.  Other
colour excesses are quite stable; $E(V-R)$, for example, has a maximum
deviation of 0.095~mag and scatter of 0.025~mag from $t = -10$ to
+74~days.  For most colour excesses, there is a measurement error of
only a few hundredths of a magnitude.

Since SN~Ia SEDs change dramatically with time and heavily reddened
(or scattered) SEDs will shift the effective wavelength of a filter,
comparisons are necessary to interpret the colour curves and transform
$E(X-Y)^{\rm obs}$ to $E(X-Y)^{\rm true}$.  Observed SN~Ia
colour-excess curves change with time \citep[e.g.,][]{Phillips99,
  Jha07}, and high-cadence measurements of the observed colour-excess
curves are necessary for proper comparison.  The \citet{Hsiao07}
template spectra were used for this purpose.  These spectra were
generated from many different SNe and are close to an ``average''
spectrum.  The high cadence of the template spectra is particularly
attractive for comparisons.  We also use the {\it HST} (and NIR when
available) spectra of SN~2011fe for validation of the template
spectra.  It is known that the \citet{Hsiao07} template spectra do not
properly describe the early-time UV behaviour of SNe~Ia
\citep{Foley12:09ig}, and the SN~2011fe data should be particularly
useful in this regime.

Model colour-excess curves are created by comparing the colour curves
of the synthetic photometry of the reddened and/or scattered template
and SN~2011fe spectral series relative to the colour curves of the
synthetic photometry from the unchanged spectra.  If SN~2014J has the
same SEDs and temporal evolution as the template spectra, the true
reddening parameters will produce colour-excess curves that match
those of SN~2014J.

The \citet{Tsvetkov14} SN~2014J $B-V$ colour curve is \about 0.05~mag
redder at peak than the one presented here.  This is likely caused by
$S$ corrections \citep{Stritzinger02}.  Such differences may produce
slight shifts (of order 0.05~mag) in the measured colour excess.
Because we have multiple bands, and these shifts should be roughly
random, the total uncertainty in $E(B-V)$ should be $<$0.01~mag.

We test the 6 reddening/scattering models listed in
Section~\ref{s:models} by reddening/scattering the template spectra
and deriving the observed colour excesses.  These measurements are
compared to the data, and best-fitting parameters for each model are
measured.  Although our primary comparison is SN~2011fe, we also test
how our results vary when using SNe~2009ig and 2013dy, both of which
have fewer data than SN~2011fe.  All fit parameters have an
uncertainty related to the unknown intrinsic colour of SN~2014J, which
is \about 0.1~mag for $E(B-V)$ and \about 0.05 for $R_{V}$ and $p$.
For some model comparisons, this additional uncertainty cancels out,
and thus we do not list it below to make more precise comparisons.

The SN~2014J data are inconsistent with the most basic reddening
models where $R_{V}$ is set to 3.1.  For the F9931 and CCM31 models,
we find best-fitting reddening values of $E(B-V)^{\rm true} = 0.874
\pm 0.018$ and $0.887 \pm 0.015$~mag with reduced $\chi^{2}$ of
$\chi^{2}_{\nu} = 10.4$ and 10.2, respectively.  These models result
in $A_{V}^{\rm true} \approx 2.7$~mag, which is much larger than our
observed and model-independent value (Section~\ref{s:ext}).  We do not
include the best-fitting colour-excess curves for these models in
Figure~\ref{f:colex}; because of their poor fit, doing so would change
the presentation of the figure to the point where one cannot
discriminate from other, better-fitting models.

In fact, all simple reddening models with $R_{V} > 2$ are not good
fits to the data.  Specifically, $R_{V} = 2$ yields $\chi^{2}_{\nu} =
3.3$ and 2.9 for the F99 and CCM models, respectively.

When allowing $R_{V}$ to be free, the best-fitting values for the F99
and CCM models are $E(B-V)^{\rm true} = 1.194 \pm 0.012$ and $1.244
\pm 0.010$~mag with $R_{V}^{\rm true} = 1.66 \pm 0.03$ and $1.44 \pm
0.03$, respectively.  These models have $\chi^{2}_{\nu} = 2.6$ and
$1.5$, respectively, indicating moderate success at reproducing the
observations.  At these low values of $R_{V}$, the derived reddening
laws are extrapolations, implying an additional systematic uncertainty
not included in this analysis.  These models imply $A_{V}^{\rm true} =
2.0$ and 1.8~mag, respectively, both similar to our model-independent
measurements of the extinction: $A_{V}^{\rm obs} = 1.95 \pm 0.09$~mag.

The circumstellar material scattering model (CSM) is significantly
better at describing the observed data.  We find a best-fitting
power-law index of $p = -1.977 \pm 0.014$, similar to LMC and Milky
Way dust, and $\chi^{2}_{\nu} = 0.54$.  Fitting the data with a simple
power law ($A_{\lambda}/A_{V} \propto (\lambda/\lambda_{V})^{p}$)
results in an equally good fit as the \citetalias{Goobar08} model with
the same power-law index.

Because of the statistical success of the CSM model, the two-component
CSMD model is not strictly necessary.  None the less, we test certain
physically motivated models and report the results.  Fixing $R_{V} =
3.1$, we find $E(B-V)^{\rm true} = 0.49 \pm 0.02$~mag and $p = -3.42
\pm 0.15$ with $\chi^{2}_{\nu} = 0.55$ --- essentially the same as for
the best-fitting CSM model.  For the Milky Way scattering parameters,
we find a worse fit, $E(B-V)^{\rm true} = 0.796 \pm 0.009$~mag and
$R_{V}^{\rm true} = 1.40 \pm 0.03$ with $\chi^{2}_{\nu} = 1.60$.  For
the LMC scattering parameters, the best-fitting parameters are
$E(B-V)^{\rm true} = 0.492 \pm 0.007$~mag and $R_{V}^{\rm true} = 2.35
\pm 0.05$ with $\chi^{2}_{\nu} = 0.83$.  Restricting the scattering
power law to $-1.5 \ge p \ge -2.5$ and $R_{V} = 3.1$, we find a
best-fitting result with $p = -2.5$ and $E(B-V)^{\rm true} = 0.273 \pm
0.008$~mag with $\chi^{2}_{\nu} = 0.58$.  Finally, using the
best-fitting parameters when fitting to the spectra
(Section~\ref{s:spec_red}), we measure $\chi^{2}_{\nu} = 0.75$.

There is significant temporal evolution in some colour excesses, which
is not matched by the evolution of the best-fitting reddened/scattered
templates.  Examining the \citet{Tsvetkov14} colour curves, which have
similar shapes to those presented here, this evolution is not the
result of $S$ corrections related to the SN~2014J photometry.
Alternatively, SN~2014J may have a different colour evolution than
SN~2011fe or the \citet{Hsiao07} templates.  Examining 29 high-quality
SN~Ia light curves \citep{Hicken09:lc, Contreras10} that cover both a
week before until a week after maximum brightness, two (7\%) have a
$B-V$ colour evolution as extreme as that of SN~2014J.  The exact
reddening parameters have a small effect on the evolution of the
comparison colour curves, which in turn affects the amount of
variability (at the \about 20\% level).  Because of these different
effects, we are not confident that there is temporal evolution of
$E(B-V)^{\rm true}$ for SN~2014J.


\section{Spectroscopic Reddening Estimates}\label{s:spec_red}

One of the best ways to determine the reddening for an astrophysical
object is to compare the SEDs of the reddened object to an identical
unreddened object.  The difference between the measured flux (after
correcting for any potential distance difference) at a given
wavelength is the extinction for that wavelength.  The extinction as a
function of wavelength, the extinction curve, can then be compared to
the extinction curves expected for different reddening laws.

Here we take a slightly different approach.  Instead of assuming the
known luminosity of the SN, we only assume that the spectral shapes of
SN~2014J and the comparison SNe are similar, leaving the differences
in luminosity and distance as a free parameter.  Consequently,
distance errors do not effect our measurements.

SNe~2009ig, 2011fe, and 2013dy are used as comparison SNe.  As
mentioned in Section~\ref{ss:comp}, these are the three SNe~Ia with
reasonable UV time series.  Table~\ref{t:phase} indicates the phase of
each SN~2014J spectrum and the comparison spectra of the other SNe.

Spectroscopy provides two particular advantages over photometry.
First, because of the reddening, the nominally UV {\it HST} filters
really probe optical wavelengths (Section~\ref{ss:filt}) and have
uncertain effective wavelengths and Milky Way extinctions.  Second,
particular spectral features (especially Ca H\&K, which is at the blue
end of the optical window) even vary significantly between SNe with
the same light-curve shape.  These features can be excluded from
spectral comparisons, but are unavoidable with photometry (unless one
completely excludes particular filters).

Because of the effective wavelengths of the photometric filters, our
{\it HST}/STIS spectra are the only proper UV data for SN~2014J.

\subsection{Method}

We tested the 6 models presented in Section~\ref{s:models} by
reddening/scattering the SN~2014J spectra to match the phase-matched
comparison spectra.  First, we fit only UV-optical data
(2500--10,000~\AA).  All SN~2014J and comparison spectra have data
over this range.  Although the SN~2014J {\it HST} spectra have data
below 2500~\AA, the flux is consistent with zero (with large
uncertainties).  We therefore ignore these data, although they may be
important for exotic models.  Some comparison spectra also have NIR
data; we fit the spectra with and without these data.  When fitting
the pre- and near-maximum brightness data, we ignore the spectral
regions covering Ca H\&K, the Ca NIR triplet, and the \ion{Si}{II}
$\lambda 6355$ feature.  Differences in line depths and velocity can
significantly affect the flux at these wavelengths in ways that are
unrelated to dust reddening.

For presentation purposes, we generally plot the pseudo-extinction
curve,
\begin{equation}
  X_{\lambda}/X_{V} = \frac{f_{\lambda}^{\rm SN} / f_{\lambda}^{\rm 14J}}{f_{\lambda}^{\rm SN} ({\rm 5500~\AA}) / f_{\lambda}^{\rm 14J} ({\rm 5500~\AA})},
\end{equation}
where $f_{\lambda}^{\rm 14J}$ and $f_{\lambda}^{\rm SN}$ are the flux
of SN~2014J and the comparison SN, respectively.  This function is
scaled by the flux ratios at the $V$ band (roughly 5500~\AA).  This
curve is not an extinction curve since the normalisation is arbitrary
to compensate for the uncertainty in the SN luminosities and
distances.  Reddening curves can then be directly compared.  When
comparing two sets of data or models to the data, we can then use the
residuals to determine $\Delta A_{\lambda}$ and still avoid
distance/luminosity uncertainties.

\subsection{Single-Component Reddening Models}\label{ss:single}

We first examine the single-component reddening models: CCM31, F9931,
CCM, and F99.  These models require a single value for $E(B-V)^{\rm
  true}$ and $R_{V}^{\rm true}$ for each spectral pair and are the
most simplistic reddening models of our tested possibilities.

\begin{figure}
\begin{center}
\includegraphics[angle=0,width=3.2in]{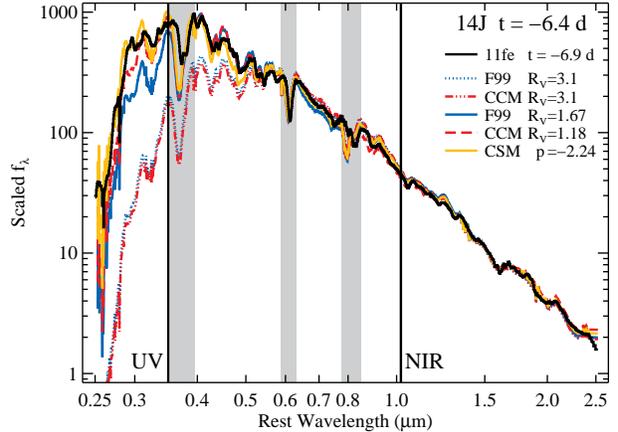}
\caption{UVOIR $t = -6.4$~day spectrum of SN~2014J dereddened or
  descattered by the best-fitting reddening (F9931, CCM31, F99, and
  CCM) and circumstellar scattering (CSM) curves to match the $t =
  -6.9$~day SN~2011fe spectrum (black curve).  We do not plot the
  best-fitting two-component circumstellar scattering and reddening
  model (CSMD) since it is very similar to the best-fitting CSM model
  (despite having different parameters).  The regions excluded from
  the fitting, corresponding to Ca H\&K, \ion{Si}{II} $\lambda 6355$,
  and the Ca NIR triplet, have been marked by the grey regions.  The
  UV and NIR regions are noted.  Most of the discriminating power
  comes from the UV, which is not probed by the photometry, but the
  NIR is helpful in anchoring the models.}\label{f:11fe1}
\end{center}
\end{figure}

The $t = -6.4$~day spectrum of SN~2014J, dereddened by the
best-fitting extinction curves, is displayed in Figure~\ref{f:11fe1}.
This figure represents the output of the spectral matching.

\begin{figure}
\begin{center}
\includegraphics[angle=0,width=3.2in]{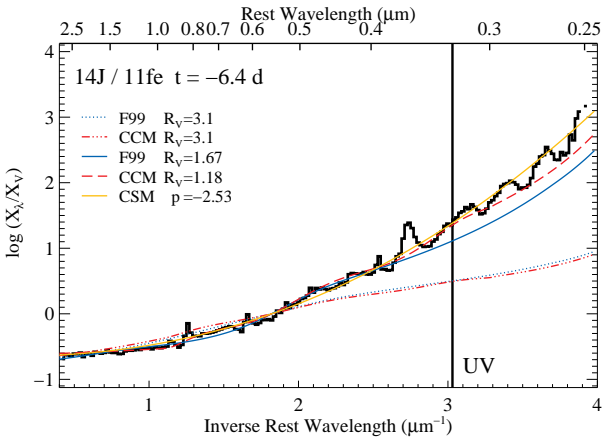}
\caption{Pseudo-extinction curve of SN~2014J at $t = -6.4$~days
  determined by comparing to SN~2011fe (black curve).  Various
  reddening curves and scattering models corresponding to those
  presented in Figure~\ref{f:11fe1} are overplotted.  The data are
  incompatible with an $R_{V} = 3.1$ reddening
  law.}\label{f:red_11fe1}
\end{center}
\end{figure}

\begin{figure}
\begin{center}
\includegraphics[angle=0,width=3.2in]{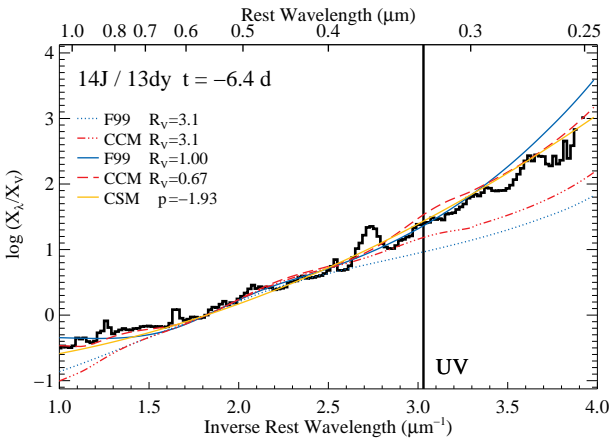}
\caption{Same as Figure~\ref{f:red_11fe1}, but using SN~2013dy as the
  comparison and with slightly different model
  parameters.}\label{f:red_13dy1}
\end{center}
\end{figure}

Figures~\ref{f:red_11fe1} and \ref{f:red_13dy1} show the
pseudo-extinction curves for the $t = -6.4$~day SN~2014J spectrum
(compared to the $t = -6.9$~day SN~2011fe and $t = -6.2$~day SN~2013dy
spectra, respectively).  Overplotted are the best-fitting extinction
curves for the above scenarios.

The most obvious result of the above analysis is that the SN~2014J
spectra are inconsistent with simple reddening with $R_{V} = 3.1$,
similar to what was determined from the photometry
(Section~\ref{ss:colex}).  Such reddening laws consistently
undercorrect the UV.

When $R_{V}$ is allowed to vary, the data are reasonably fit with a
single reddening law; however, extremely low values for $R_{V}$ are
required.  For instance, using the \citetalias{Cardelli89} law and
comparing to SN~2013dy, the highest best-fitting value for any
spectral pair is $R_{V} = 2.0$.

\begin{figure}
\begin{center}
\includegraphics[angle=0,width=3.2in]{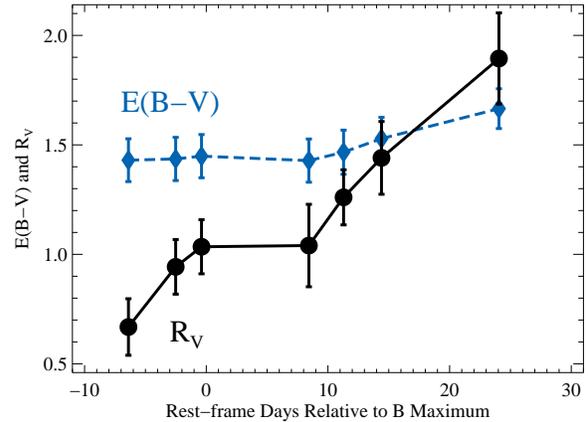}
\caption{Measurements of $E(B-V)^{\rm true}$ (blue diamonds) and
  $R_{V}^{\rm true}$ (black circles) for SN~2014J as a function of
  time.  The measurements are made by comparing spectra of SN~2014J to
  those of SN~2013dy at similar phases.  We assume a
  \citetalias{Cardelli89} reddening law, and SN~2013dy has been
  dereddened by $E(B-V)_{\rm host} = 0.15$~mag.  The uncertainties are
  dominated by the uncertainty in the host-galaxy reddening of
  SN~2013dy.}\label{f:rv_evol}
\end{center}
\end{figure}

The best-fitting values for the CCM and F99 models change dramatically
with time.  Figure~\ref{f:rv_evol} displays the best-fitting CCM
values of $E(B-V)^{\rm true}$ and $R_{V}^{\rm true}$ for SN~2014J as a
function of time.  Each SN~2014J spectrum is fit separately.  All
epochs are consistent with $E(B-V)^{\rm true} = 1.49 \pm 0.09$~mag,
but $R_{V}^{\rm true}$ increases from 0.67 to 1.89 over the span of
one month.  The UV portion of our spectra is not the cause of this
change; using only data with $\lambda > 3000$~\AA, we find a mean
reddening of $E(B-V)^{\rm true} = 1.55 \pm 0.10$~mag and $R_{V}^{\rm
  true}$ increasing from 0.76 to 1.65.  The photometry of SN~2014J
showed similar behaviour with $E(B-V)^{\rm obs}$, although we were not
confident that the photometric derivation of $E(B-V)^{\rm obs}$ was
truly varying.

No physical model of dust formation/destruction or different clouds of
dust entering/exiting the SN beam can have unchanging reddening and
changing $R_{V}$.  It is possible that dust grains coagulate to form
larger grains on average, but it is unlikely that this process would
occur on these timescales.  Therefore, these simple reddening models
cannot explain the SN~2014J observations.

Many of the best-fitting values for $R_{V}$ are unphysically low.  The
expectation for Rayleigh scattering is $R_{V} \approx 1.2$, with
smaller values being extremely unlikely.  This alone may indicate a
problem with a pure reddening scenario; however, the reddening laws
are extrapolations for these values of $R_{V}$, and thus the true
value of $R_{V}$ may be larger than what is measured.

\subsection{Circumstellar Scattering}\label{ss:csm}

Using the CSM scattering model described in Section~\ref{s:models}, we
can correct the SN~2014J spectra to be roughly consistent with the
spectra of the other SNe.  However, circumstellar scattering does not
provide a better fit than the reddening laws (over all spectra).

Similar to the dust reddening parameters, the circumstellar scattering
parameters change with epoch.  For instance, the power-law exponent,
$p$, changes from $-2.1$ to $-1.4$ when comparing to SN~2013dy.
Although this is potentially physically possible if dust were being
actively destroyed, revealing a different grain-size distribution, no
specific prediction has been made.

\subsection{Variable Extinction?}

\begin{figure}
\begin{center}
\includegraphics[angle=0,width=3.2in]{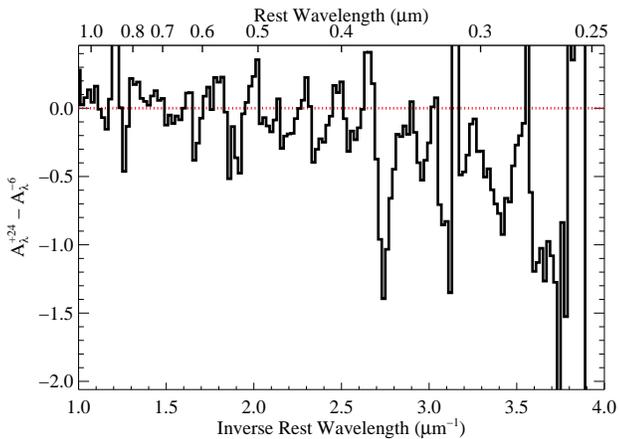}
\caption{Difference in the extinction curves for SN~2014J at $t = -6$
  and $+24$~days as determined by comparing to SN~2011fe spectra at
  similar epochs.}\label{f:ext_comp}
\end{center}
\end{figure}

Given that the best-fitting parameters change as SN~2014J evolves with
time, we wish to determine if the changing parameters are caused by
physical differences or poor fitting.  Using SN~2011fe as an
intermediary, we compare the extinction curves of SN~2014J at $t =
-6.4$~days and $t = +24.1$~days in Figure~\ref{f:ext_comp}.  For the
two epochs, the extinction is similar in the optical, but differs
significantly in the UV with the $t = -6.4$~day extinction curve
having significantly more extinction in the UV.

Either the extinction curves are changing with time or SN~2014J has
significantly different colour evolution than SN~2011fe.  SN~2014J is
spectrally most similar to SN~2007co, which has a similar colour
evolution to SN~2011fe, but is redder for all epochs.  Therefore,
SN~2011fe should be a good comparison for this task, although the
exact extinction difference at a given wavelength may be slightly
different if SN~2007co were used\footnote{Unfortunately, there are no
  existing UV spectra of SN~2007co, and thus a comparison at the most
  interesting wavelengths cannot be made.}.

\citet{Patat06} showed that for circumstellar dust the inferred
extinction curve, when comparing spectral pairs, should change
dramatically from maximum until at least one month after maximum
brightness.  The derived curves should be roughly the same in the red
for both epochs, but in the blue, the later spectral comparison should
indicate a decrease in $A_{\lambda}$ relative to what was measured at
maximum brightness.  SN~2014J displays this behaviour in
Figure~\ref{f:ext_comp}.

\subsection{Multiple-Component Extinction}

Given the deficiencies in single-component models describing the
extinction toward SN~2014J, we turn to the CSMD model.  Given the
complexity of M82 and the fact that SN~2014J is in its disc, there is
likely ISM dust in addition to any possible circumstellar dust or
scattering.

Adding a dust component to the circumstellar scattering model, CSMD,
resulted in two components of roughly equal weight.  These models have
best-fitting $E(B-V) \approx 0.6$~mag, roughly half that of the
single-component models.  Intriguingly, $R_{V}$ also generally
increased.  Meanwhile, the parameters for the scatter component were
$a \approx 0.6$ and $p \approx -2.7$, roughly consistent with the LMC
dust model of \citetalias{Goobar08} and similar to good fits to the
photometry (Section~\ref{ss:colex}).  However, we caution that there
are some degeneracies between these components, and the solutions are
not particularly well determined for any specific spectral pair.  To
compensate for these degeneracies, we also fit the entire time series
at once.

\begin{figure}
\begin{center}
\includegraphics[angle=0,width=3.2in]{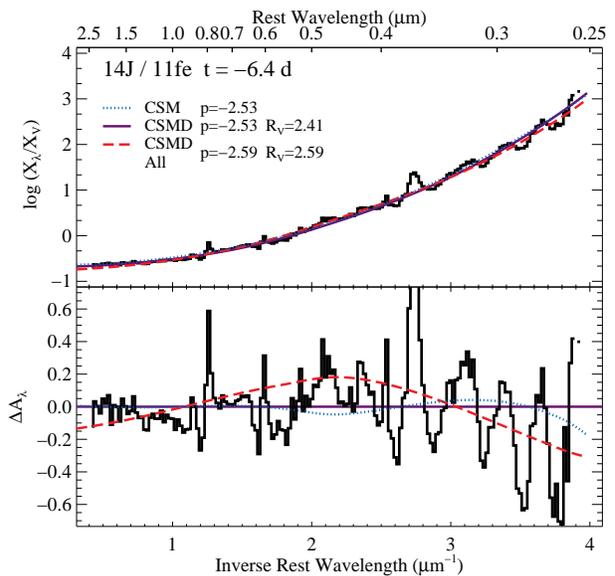}
\caption{({\it Top panel}): Pseudo-extinction curve of SN~2014J at $t
  = -6.4$~days determined by comparing to SN~2011fe (black curve).
  Various reddening curves and scattering models are overplotted.
  Similar to Figure~\ref{f:red_11fe1}.  ({\it Bottom panel}): Residual
  extinction of SN~2014J at $t = -6.4$~days relative to the
  best-fitting two-component dust reddening and circumstellar
  scattering model.  Also shown are the residual extinction of
  two-component dust reddening and circumstellar scattering models
  with best-fitting parameters for the full spectral time series
  (dashed red curve) and the circumstellar scattering only model for
  this epoch (dotted blue curve).}\label{f:red2_11fe1}
\end{center}
\end{figure}

\begin{figure}
\begin{center}
\includegraphics[angle=0,width=3.2in]{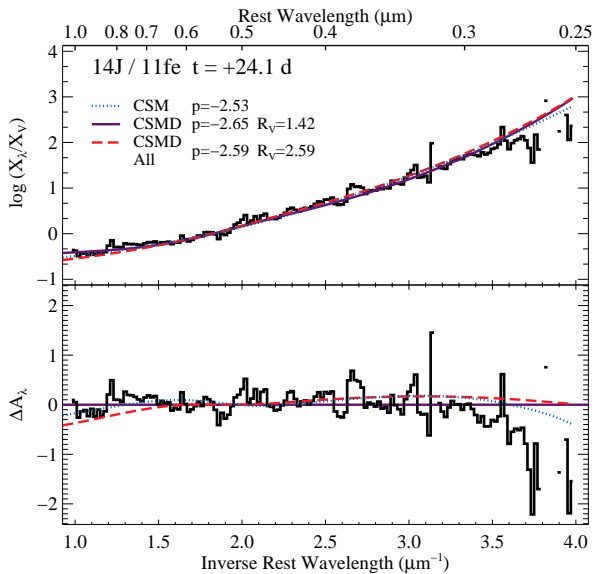}
\caption{Same as Figure~\ref{f:red2_11fe1}, but for the $t = 24.1$~day
  spectrum of SN~2014J.}\label{f:red2_11fe10}
\end{center}
\end{figure}

The pseudo-extinction curves for SN~2014J at $t = -6.4$ and 24.1~days
are presented in Figures~\ref{f:red2_11fe1} and \ref{f:red2_11fe10},
respectively.  They are compared to the best-fitting CSM and CSMD
models for each individual epoch and the best-fitting CSMD model for
all epochs.  For these two examples, the various models do not vary
significantly.

While the single set of best-fitting parameters for all spectra does
not provide the best fit for any individual spectrum (as one would
expect), the $\chi^{2}_{\nu}$ for each spectrum does not dramatically
increase.  The best-fitting values for this case (when comparing to
SN~2011fe) are $E(B-V)^{\rm true} = 0.45 \pm 0.02$~mag, $R_{V}^{\rm
  true} = 2.59 \pm 0.02$, $a = 0.83 \pm 0.05$, and $p = -2.60 \pm
0.06$ for a \citetalias{Fitzpatrick99} reddening law.  These values
represent a reasonable amount of dust reddening with typical
properties and circumstellar scattering off LMC-like dust.

We also show the CSMD colour-excess curves for these parameters in
Figure~\ref{f:colex}.  This model is a good fit to the photometric
data in addition to the spectral data.

\begin{figure}
\begin{center}
\includegraphics[angle=0,width=3.2in]{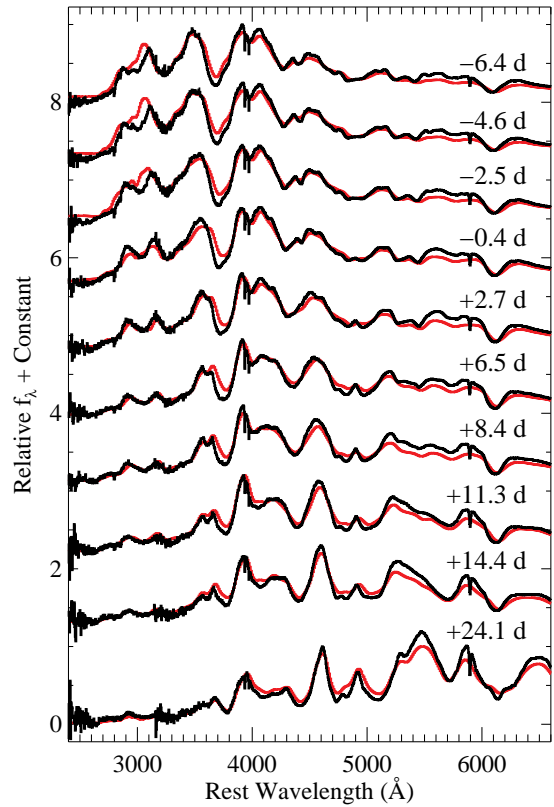}
\caption{Our full {\it HST} spectral series of SN~2014J (black curves)
  dereddened by a \citetalias{Fitzpatrick99} reddening law with
  $E(B-V) = 0.45$~mag and $R_{V} = 2.57$, and descattered by a
  \citetalias{Goobar08} circumstellar scattering model with $a = 0.84$
  and $p = -2.59$ (the globally best-fitting parameters for the CSMD
  model).  Template spectra \citep{Hsiao07} at similar epochs are
  shown in red.  The pre-maximum brightness template spectra are known
  to be inaccurate at UV wavelengths.}\label{f:csmd}
\end{center}
\end{figure}

We present our SN~2014J {\it HST} spectra, dereddened and descattered
by the globally best-fitting CSMD parameters, in Figure~\ref{f:csmd}.
The spectra are compared to the \citet{Hsiao07} template spectra at
similar epochs.  Overall there is good agreement, especially
considering that the parameters are not the result of fitting to the
template spectra.

\subsection{Model Comparison}

The SN~2014J spectroscopy is inconsistent with pure reddening laws
with $R_{V} > 2$.  The three reasonable models based on spectral
comparisons alone are a dust-reddening law with $R_{V} < 2$, the CSM
model, and the hybrid CSMD model.

Allowing reddening parameters to vary for each spectral pair, we find
a total (over all spectra) $\chi^{2}_{\nu} = 37.1$, 27.6, and 23.2 for
the F99, CSM, and CSMD models, respectively.  These relatively large
values result from spectral features causing the SN~2014J reddening
law to deviate from smooth curves; relative comparisons are still
indicative of the quality of fitting.  The CSM and CSMD models perform
better than the F99 model.  Since the F99 model is an extrapolation to
$R_{V} < 2$, the poor fitting may be the result of this extrapolation.

Forcing all epochs to have the same reddening parameters, we find a
total $\chi^{2}_{\nu} = 72.1$, 53.4, and 50.3 for the F99, CSM, and
CSMD models, respectively.  The larger $\chi^{2}_{\nu}$ values are
caused by the changing continuum with time, which cannot be accounted
for by fixing parameters for all epochs.  All $\chi^{2}_{\nu}$ values
doubled by fixing these parameters.  Again, the CSM and CSMD models
perform better than the F99 model.


\section{Discussion \& Conclusions}\label{s:disc}

We presented 10 {\it HST}/STIS spectra of SN~2014J from $-6.4$ to
+24.1~days relative to $B$-band maximum brightness.  We supplemented
these UV-optical spectra with several NIR spectra of SN~2014J.  We
also present 17-band photometry from {\it HST}/WFC3, KAIT, and FanCam.
Because of red leaks in the WFC3 UV filters, our photometry does not
effectively probe the UV SED of SN~2014J.  This extensive dataset is
one of the best for a SN~Ia in terms of temporal and wavelength
sampling.

There is significant extinction toward SN~2014J.  Previous studies
have applied inaccurate Milky Way reddening corrections to SN~2014J,
which have biased measurements of the SN~2014J extinction arising in
M82.  Using model-independent methods, we determine that extinction to
be $A_{V}^{\rm obs} = 1.95 \pm 0.09$~mag and the reddening to be
$E(B-V)^{\rm obs} = 1.19 \pm 0.14$~mag.  Comparing these values, we
find $R_{V}^{\rm obs} = 1.64 \pm 0.16$.

Because of the complex dust structures in M82 and the position of
SN~2014J within M82, it is extremely likely that there is some
interstellar dust reddening in M82.  SN~2014J is offset by a projected
distance of 1.0~kpc from the nucleus of M82, located in the disc
(Figure~\ref{f:finder}).  \citet{Hutton14} examined the dust
properties of M82, finding a luminosity-weighted reddening of $E(B-V)
\approx 0.5$~mag for a projected distance of 1~kpc with $R_{V} \approx
3.1$.

Strong DIB absorption in high-resolution spectra of SN~2014J also
indicates that there is a relatively large amount of dust reddening.
For the 5780~\AA\ DIB feature, we measured an equivalent width of $344
\pm 21$~m\AA\ (including a Milky Way contribution which should be
$\lesssim$5\% of the total).  This value is consistent with that found
by \citet{Welty14}, but highly inconsistent with that of
\citet{Goobar14}.  The DIB measurement implies $A_{V} = 1.8 \pm
0.9$~mag (or $A_{V} = 2.5 \pm 1.3$~mag for the \citealt{Goobar14}
measurement).

Examining the colour excesses in multiple bands and the UV through NIR
spectra, which do not depend on the distance to M82, we determine that
dust with $R_{V} = 3.1$, the canonical value for the Milky Way, cannot
be the exclusive source of reddening for SN~2014J.  In fact, we place
a strong constraint of $R_{V} < 2$ for single-component dust
reddening.

From our measured colour excesses, the best-fitting parameters for the
F99 and CCM models are $E(B-V)^{\rm true} = 1.19 \pm 0.10$ and $1.24
\pm 0.10$~mag with $R_{V} = 1.66 \pm 0.06$ and $1.44 \pm 0.06$
(including systematic uncertainties), respectively, from the
photometry.  Reddening the maximum-brightness SN~2011fe spectrum by
these values, we measure $A_{V}^{\rm obs} = 1.91$~mag (1.82~mag) and
$E(B-V)^{\rm obs} = 1.19$~mag (1.17~mag) for the F99 (CCM) model,
consistent with the model-independent measurements.

The best-fitting $R_{V}$ for SN~2014J is consistent with those of
other highly reddened SNe~Ia \citep[e.g.,][]{Elias-Rosa06,
  Elias-Rosa08, Krisciunas06, Wang08:06x}, but are inconsistent with
essentially all local lines of sight in the Milky Way
\citep[e.g.,][]{Fitzpatrick07}.  The extinction measured from the
spectroscopy varies with epoch, which is difficult to explain with
only ISM dust reddening.  Using a model-independent approach, we see
that the extinction at shorter wavelengths increases with time.
Because of the variations seen in the spectra, systematic effects
related to the effective wavelength of the photometric filters or
differing spectral features between SN~2014J and comparison SNe are
not the cause of the variation.

Using the same {\it HST} photometric data presented here (although
independently reduced) combined with independent optical and NIR
photometry, \citet{Amanullah14} examined the reddening toward
SN~2014J, concluding that $R_{V} \lesssim 2$ for dust-reddening
scenarios.  \citet{Amanullah14} suggest best-fitting parameters of
$E(B-V)^{\rm true} = 1.37 \pm 0.03$ and $R_{V}^{\rm true} = 1.4 \pm
0.1$ for a \citetalias{Fitzpatrick99} reddening law.  \citet{Goobar14}
used the \citet{Schlafly11} catalogue value of $E(B-V) = 0.14$~mag for
the Milky Way reddening.  \citet{Amanullah14} used the raw
\citet{Dalcanton09} value of $E(B-V) = 0.06$~mag for the Milky Way
reddening, but this value is on the \citet{Schlegel98} scale.  We use
the \citet{Dalcanton09} value on the \citet{Schlafly11} scale: $E(B-V)
= 0.05$~mag.  Adjusting the \citet{Amanullah14} reddening value to
account for the correct Milky Way extinction results in $E(B-V)^{\rm
  true} = 1.38$~mag.  This value is larger than our best-fitting
value, but is consistent.  Reddening the maximum-brightness SN~2011fe
spectrum by the corrected values, we measure $A_{V}^{\rm obs} =
1.84$~mag and $E(B-V)^{\rm obs} = 1.39$~mag for a
\citetalias{Fitzpatrick99} reddening law, which are consistent with
the model-independent measurements.

A relatively simple circumstellar scattering model
\citepalias{Goobar08} can reproduce the reddening of SN~2014J.  The
best-fitting CSM model for our photometry has $p = -1.98 \pm 0.05$,
consistent with the \citet{Amanullah14} value of $p = -2.1 \pm 0.1$.

Variable extinction and colour excess is predicted in circumstellar
matter scattering models \citep{Wang05, Patat06}.  The
\citetalias{Goobar08} model does not attempt to model the temporal
variability that must be introduced through the scattering process.
None the less, some specific predictions of \citet{Patat06} are well
matched by SN~2014J.

From about a week before maximum brightness until about a month after,
narrow ISM and/or CSM absorption features in the spectra of SN~2014J
do not vary.  Previous detections of variability have been interpreted
as evidence for CSM \citep{Patat07:06x, Blondin09, Simon09}.  A lack
of variability may constrain the mass and distance to any potential
CSM.  Our current observations cannot rule out the existence of CSM,
but are consistent with all absorbing gas being purely interstellar.

Given the position of SN~2014J in M82 and the strong DIB absorption,
it is incredibly unlikely that there is {\it no} interstellar dust
reddening for SN~2014J.  Therefore, the CSM model is unlikely from
this simple argument.  The extreme values for $R_{V}$ and the variable
extinction also argue against a simple dust-reddening model.  Although
neither model is ruled out by our data, a promising model is a
combination of the two.

Fitting a two-component circumstellar scattering and dust reddening
model to the data, we find a consistent picture.  The reddening of
SN~2014J is well reproduced by a reasonable amount of reddening
($E(B-V) = 0.45$~mag) from relatively typical dust ($R_{V} = 2.59$)
with circumstellar dust parameters similar to those for LMC-like dust.
These values represent the best measurements of the dust and reddening
properties for SN~2014J.  In this scenario, roughly half of the
extinction is caused by dust reddening ($A_{V} = 1.17$~mag) and the
other half from scattering.  The dust-reddening parameters for this
model are remarkably similar to the estimates of the reddening based
on the position of SN~2014J \citep{Hutton14}.  They are also
consistent with the implied reddening from the high-resolution
spectroscopy.  Finally, circumstellar scattering predicts a change in
the measured extinction at bluer wavelengths with time, consistent
with our observations.

It is unlikely that SN~2014J is the only SN~Ia with both dust
reddening and circumstellar scattering.  Re-examination of SNe~Ia with
high reddening and measurements of a low $R_{V}$ may reveal similar
configurations for these SNe as well.  The best-fitting value of
$R_{V}$ for SN~2014J is 2.59, similar to many lines of sight in the
Milky Way and similar to that found for low-reddening SNe~Ia (when
properly accounting for intrinsic colour).  Therefore, a CSMD model
may solve the problem of small measured values of $R_{V}$ for
high-reddening SNe~Ia.

The {\it HST} UV spectra are very constraining for the reddening
models.  Thankfully, SNe~2011fe and 2013dy had 10 epochs of
spectroscopy each, allowing for several comparisons of spectra with
similar phases.  Additional SNe~Ia with UV spectroscopy will further
improve constraints for SN~2014J and future events.

More detailed modelling of how circumstellar scattering affects the
SED with time may be the best way to constrain CSM parameters using SN
SEDs.  However, more direct methods such as observations of the
circumstellar gas in absorption \citep[e.g.,][]{Patat07:06x,
  Sternberg11, Foley12:csm, Maguire13} and radio/X-ray observations
\citep[e.g.,][]{Chomiuk12, Horesh12, Margutti12, Margutti14} may
provide the best understanding of the circumstellar environments of
SNe~Ia.

Being the closest detected SN~Ia in at least 28 years, and perhaps in
410 years, SN~2014J will be an important event for understanding SN~Ia
physics for decades.  Here, we have begun to unravel the reddening, a
key component in further understanding of the event.  A proper
measurement of the dust reddening will be especially important for
multi-wavelength studies and the long-term monitoring of SN~2014J.

\section*{Acknowledgements}

Based on observations made with the NASA/ESA Hubble Space Telescope,
obtained from the Data Archive at the Space Telescope Science
Institute, which is operated by the Association of Universities for
Research in Astronomy, Inc., under NASA contract NAS 5--26555. These
observations are associated with programs GO--13286 and DD--13621.  We
thank the Director for observing SN~2014J through program DD--13621.
We especially thank the STScI staff for accommodating our
target-of-opportunity program.  A.\ Armstrong, R.\ Bohlin, S.\
Holland, S.\ Meyett, and D.\ Taylor were critical for the execution of
this program.

We thank C.\ Contreras, S.\ Taubenberger, and C.\ Wheeler for helpful
discussions; I.\ Arcavi, M.\ Graham, A.\ Howell, J.\ Parrent, and S.\
Valenti for their support with the IRTF observations; and P.\ Berlind,
L.\ Buchhave, M.\ Calkins, G.\ Esquerdo, G.\ Furesz, and A.\
Szentgyorgi for help with the TRES observations.  We especially thank
D.\ Latham for proving target-of-opportunity time with TRES.  We
appreciate the efforts of J.\ Tobin and the undergraduate
Observational Astronomy class at the University of Virginia, who
supplied the observations for three epochs of the NIR photometry.

This manuscript was completed during the ``Fast and Furious:
Understanding Exotic Astrophysical Transients'' workshop at the Aspen
Center for Physics, which is supported in part by the NSF under grant
No.\ PHYS-1066293.  R.J.F., O.D.F., A.V.F., R.P.K., and J.M.S.\ thank
the Aspen Center for Physics for its hospitality during the ``Fast and
Furious'' workshop in June 2014.

Supernova research at Harvard is supported in part by NSF grant
AST--1211196.
G.H.M.\ and D.J.S.\ are visiting Astronomers at the Infrared Telescope
Facility, which is operated by the University of Hawaii under
Cooperative Agreement no.\ NNX--08AE38A with the National Aeronautics
and Space Administration.
G.P.\ acknowledges support by the Ministry of Economy, Development,
and Tourism's Millennium Science Initiative through grant IC12009,
awarded to The Millennium Institute of Astrophysics, MAS.
F.K.R.\ was supported by the Emmy Noether Program (RO 3676/1-1) of the
Deutsche Forschungsgemeinschaft and the ARCHES prize of the German
Ministry of Education and Research (BMBF).
J.M.S.\ is supported by an NSF Astronomy and Astrophysics Postdoctoral
Fellowship under award AST--1302771.
I.R.S.\ is supported by the ARC Laureate Grant FL0992131.
M.D.S.\ gratefully acknowledges generous support provided by the
Danish Agency for Science and Technology and Innovation realised
through a Sapere Aude Level 2 grant.
A.V.F.'s supernova group at UC Berkeley is supported through NSF grant
AST--1211916, the TABASGO Foundation, and the Christopher R. Redlich 
Fund.
KAIT and its ongoing operation were made possible by donations 
from Sun Microsystems, Inc., the Hewlett-Packard Company, AutoScope 
Corporation, Lick Observatory, the NSF, the University of California, 
the Sylvia \& Jim Katzman Foundation, and the TABASGO Foundation.

This research has made use of the NASA/IPAC Extragalactic Database
(NED) which is operated by the Jet Propulsion Laboratory, California
Institute of Technology, under contract with the National Aeronautics
and Space Administration.

\bibliographystyle{mn2e}
\bibliography{../astro_refs}

\begin{thebibliography}{108}
\expandafter\ifx\csname natexlab\endcsname\relax\def\natexlab#1{#1}\fi

\bibitem[{{Amanullah} \& {Goobar}(2011)}]{Amanullah11}
{Amanullah} R., {Goobar} A., 2011, \apj, 735, 20

\bibitem[{{Amanullah} {et~al}\mbox{.}(2014){Amanullah}, {Goobar}, {Johansson},
  {Banerjee}, {Venkataraman}, {Joshi}, {Ashok}, {Cao}, {Kasliwal}, {Kulkarni},
  {Nugent}, {Petrushevska}, \& {Stanishev}}]{Amanullah14}
{Amanullah} R. {et~al.}, 2014, \apjl, 788, L21

\bibitem[{{Ayani}(2014)}]{Ayani14}
{Ayani} K., 2014, Central Bureau Electronic Telegrams, 3792, 1

\bibitem[{{Betoule} {et~al}\mbox{.}(2014){Betoule}, {Kessler}, {Guy}, {Mosher},
  {Hardin}, {Biswas}, {Astier}, {El-Hage}, {Konig}, {Kuhlmann}, {Marriner},
  {Pain}, {Regnault}, {Balland}, {Bassett}, {Brown}, {Campbell}, {Carlberg},
  {Cellier-Holzem}, {Cinabro}, {Conley}, {D'Andrea}, {DePoy}, {Doi}, {Ellis},
  {Fabbro}, {Filippenko}, {Foley}, {Frieman}, {Fouchez}, {Galbany}, {Goobar},
  {Gupta}, {Hill}, {Hlozek}, {Hogan}, {Hook}, {Howell}, {Jha}, {Le Guillou},
  {Leloudas}, {Lidman}, {Marshall}, {M{\"o}ller}, {Mour{\~a}o}, {Neveu},
  {Nichol}, {Olmstead}, {Palanque-Delabrouille}, {Perlmutter}, {Prieto},
  {Pritchet}, {Richmond}, {Riess}, {Ruhlmann-Kleider}, {Sako}, {Schahmaneche},
  {Schneider}, {Smith}, {Sollerman}, {Sullivan}, {Walton}, \&
  {Wheeler}}]{Betoule14}
{Betoule} M. {et~al.}, 2014, ArXiv e-prints

\bibitem[{{Blondin} {et~al}\mbox{.}(2009){Blondin}, {Prieto}, {Patat},
  {Challis}, {Hicken}, {Kirshner}, {Matheson}, \& {Modjaz}}]{Blondin09}
{Blondin} S., {Prieto} J.~L., {Patat} F., {Challis} P., {Hicken} M., {Kirshner}
  R.~P., {Matheson} T., {Modjaz} M., 2009, \apj, 693, 207

\bibitem[{{Blondin} \& {Tonry}(2007)}]{Blondin07}
{Blondin} S., {Tonry} J.~L., 2007, \apj, 666, 1024

\bibitem[{{Burns} {et~al}\mbox{.}(2011){Burns}, {Stritzinger}, {Phillips},
  {Kattner}, {Persson}, {Madore}, {Freedman}, {Boldt}, {Campillay},
  {Contreras}, {Folatelli}, {Gonzalez}, {Krzeminski}, {Morrell}, {Salgado}, \&
  {Suntzeff}}]{Burns11}
{Burns} C.~R. {et~al.}, 2011, \aj, 141, 19

\bibitem[{{Cao} {et~al}\mbox{.}(2014){Cao}, {Kasliwal}, {McKay}, \&
  {Bradley}}]{Cao14}
{Cao} Y., {Kasliwal} M.~M., {McKay} A., {Bradley} A., 2014, The Astronomer's
  Telegram, 5786, 1

\bibitem[{{Cardelli} {et~al}\mbox{.}(1989){Cardelli}, {Clayton}, \&
  {Mathis}}]{Cardelli89}
{Cardelli} J.~A., {Clayton} G.~C., {Mathis} J.~S., 1989, \apj, 345, 245

\bibitem[{{Chandler} \& {Marvil}(2014)}]{Chandler14}
{Chandler} C.~J., {Marvil} J., 2014, The Astronomer's Telegram, 5812, 1

\bibitem[{{Chandra} {et~al}\mbox{.}(2014){Chandra}, {Basu}, {Ray}, \&
  {Chakraborty}}]{Chandra14}
{Chandra} P., {Basu} A., {Ray} A., {Chakraborty} S., 2014, The Astronomer's
  Telegram, 5804, 1

\bibitem[{{Chomiuk} {et~al}\mbox{.}(2012){Chomiuk}, {Soderberg}, {Moe},
  {Chevalier}, {Rupen}, {Badenes}, {Margutti}, {Fransson}, {Fong}, \&
  {Dittmann}}]{Chomiuk12}
{Chomiuk} L. {et~al.}, 2012, \apj, 750, 164

\bibitem[{{Chomiuk} {et~al}\mbox{.}(2014){Chomiuk}, {Zauderer}, {Margutti}, \&
  {Soderberg}}]{Chomiuk14}
{Chomiuk} L., {Zauderer} B.~A., {Margutti} R., {Soderberg} A., 2014, The
  Astronomer's Telegram, 5800, 1

\bibitem[{{Contreras} {et~al}\mbox{.}(2010){Contreras}, {Hamuy}, {Phillips},
  {Folatelli}, {Suntzeff}, {Persson}, {Stritzinger}, {Boldt}, {Gonz{\'a}lez},
  {Krzeminski}, {Morrell}, {Roth}, {Salgado}, {Jos{\'e} Maureira}, {Burns},
  {Freedman}, {Madore}, {Murphy}, {Wyatt}, {Li}, \& {Filippenko}}]{Contreras10}
{Contreras} C. {et~al.}, 2010, \aj, 139, 519

\bibitem[{{Cushing} {et~al}\mbox{.}(2004){Cushing}, {Vacca}, \&
  {Rayner}}]{Cushing04}
{Cushing} M.~C., {Vacca} W.~D., {Rayner} J.~T., 2004, \pasp, 116, 362

\bibitem[{{Dalcanton} {et~al}\mbox{.}(2009){Dalcanton}, {Williams}, {Seth},
  {Dolphin}, {Holtzman}, {Rosema}, {Skillman}, {Cole}, {Girardi}, {Gogarten},
  {Karachentsev}, {Olsen}, {Weisz}, {Christensen}, {Freeman}, {Gilbert},
  {Gallart}, {Harris}, {Hodge}, {de Jong}, {Karachentseva}, {Mateo}, {Stetson},
  {Tavarez}, {Zaritsky}, {Governato}, \& {Quinn}}]{Dalcanton09}
{Dalcanton} J.~J. {et~al.}, 2009, \apjs, 183, 67

\bibitem[{{Denisenko} {et~al}\mbox{.}(2014){Denisenko}, {Gorbovskoy},
  {Lipunov}, {Balanutsa}, {Tiurina}, {Kornilov}, {Shatskiy}, {Chazov},
  {Kuznetsov}, {Rufanov}, {Vladimirov}, {Yecheistov}, {Ivanov}, {Yazev},
  {Budnev}, {Konstantinov}, {Chuvalaev}, {Poleshchuk}, {Gress}, {Yurkov},
  {Sergienko}, {Varda}, {Sinyakov}, {Gabovich}, {Parkhomenko}, {Tlatov},
  {Dormidontov}, {Senik}, {Krushinsky}, {Zalozhnih}, {Popov}, {Bourdanov},
  {Podvorotny}, {Shumkov}, {Shurpakov}, {Levato}, {Saffe}, {Mallamaci},
  {Lopez}, \& {Podest}}]{Denisenko14}
{Denisenko} D. {et~al.}, 2014, The Astronomer's Telegram, 5795, 1

\bibitem[{{Dhungana} {et~al}\mbox{.}(2014){Dhungana}, {Silverman}, {Vinko},
  {Wheeler}, {Marion}, {Kehoe}, \& {Ferrante}}]{Dhungana14}
{Dhungana} G., {Silverman} J.~M., {Vinko} J., {Wheeler} J.~C., {Marion} G.~H.,
  {Kehoe} R., {Ferrante} F.~V., 2014, Central Bureau Electronic Telegrams,
  3792, 1

\bibitem[{{Elias-Rosa} {et~al}\mbox{.}(2006){Elias-Rosa}, {Benetti},
  {Cappellaro}, {Turatto}, {Mazzali}, {Patat}, {Meikle}, {Stehle},
  {Pastorello}, {Pignata}, {Kotak}, {Harutyunyan}, {Altavilla}, {Navasardyan},
  {Qiu}, {Salvo}, \& {Hillebrandt}}]{Elias-Rosa06}
{Elias-Rosa} N. {et~al.}, 2006, \mnras, 369, 1880

\bibitem[{{Elias-Rosa} {et~al}\mbox{.}(2008){Elias-Rosa}, {Benetti}, {Turatto},
  {Cappellaro}, {Valenti}, {Arkharov}, {Beckman}, {di Paola}, {Dolci},
  {Filippenko}, {Foley}, {Krisciunas}, {Larionov}, {Li}, {Meikle},
  {Pastorello}, {Valentini}, \& {Hillebrandt}}]{Elias-Rosa08}
{Elias-Rosa} N. {et~al.}, 2008, \mnras, 384, 107

\bibitem[{{Filippenko} {et~al}\mbox{.}(2001){Filippenko}, {Li}, {Treffers}, \&
  {Modjaz}}]{Filippenko01}
{Filippenko} A.~V., {Li} W.~D., {Treffers} R.~R., {Modjaz} M., 2001, in ASP
  Conf. Ser. 246: IAU Colloq. 183: Small Telescope Astronomy on Global Scales,
  {Paczynski} B., {Chen} W.-P., {Lemme} C., eds., pp. 121--+

\bibitem[{{Fitzpatrick}(1999)}]{Fitzpatrick99}
{Fitzpatrick} E.~L., 1999, \pasp, 111, 63

\bibitem[{{Fitzpatrick} \& {Massa}(2007)}]{Fitzpatrick07}
{Fitzpatrick} E.~L., {Massa} D., 2007, \apj, 663, 320

\bibitem[{{Folatelli} {et~al}\mbox{.}(2010){Folatelli}, {Phillips}, {Burns},
  {Contreras}, {Hamuy}, {Freedman}, {Persson}, {Stritzinger}, {Suntzeff},
  {Krisciunas}, {Boldt}, {Gonz{\'a}lez}, {Krzeminski}, {Morrell}, {Roth},
  {Salgado}, {Madore}, {Murphy}, {Wyatt}, {Li}, {Filippenko}, \&
  {Miller}}]{Folatelli10}
{Folatelli} G. {et~al.}, 2010, \aj, 139, 120

\bibitem[{{Foley}(2012)}]{Foley12:vel}
{Foley} R.~J., 2012, \apj, 748, 127

\bibitem[{{Foley}(2013)}]{Foley13:ca}
{Foley} R.~J., 2013, \mnras, 435, 273

\bibitem[{{Foley}(2014)}]{Foley14:14jhst}
{Foley} R.~J., 2014, The Astronomer's Telegram, 5811, 1

\bibitem[{{Foley} {et~al}\mbox{.}(2012{\natexlab{a}}){Foley}, {Challis},
  {Filippenko}, {Ganeshalingam}, {Landsman}, {Li}, {Marion}, {Silverman},
  {Beaton}, {Bennert}, {Cenko}, {Childress}, {Guhathakurta}, {Jiang},
  {Kalirai}, {Kirshner}, {Stockton}, {Tollerud}, {Vink{\'o}}, {Wheeler}, \&
  {Woo}}]{Foley12:09ig}
{Foley} R.~J. {et~al.}, 2012{\natexlab{a}}, \apj, 744, 38

\bibitem[{{Foley} \& {Kasen}(2011)}]{Foley11:vel}
{Foley} R.~J., {Kasen} D., 2011, \apj, 729, 55

\bibitem[{{Foley} \& {Kirshner}(2013)}]{Foley13:met}
{Foley} R.~J., {Kirshner} R.~P., 2013, \apjl, 769, L1

\bibitem[{{Foley} {et~al}\mbox{.}(2012{\natexlab{b}}){Foley}, {Kromer}, {Howie
  Marion}, {Pignata}, {Stritzinger}, {Taubenberger}, {Challis}, {Filippenko},
  {Folatelli}, {Hillebrandt}, {Hsiao}, {Kirshner}, {Li}, {Morrell},
  {R{\"o}pke}, {Ciaraldi-Schoolmann}, {Seitenzahl}, {Silverman}, {Simcoe},
  {Berta}, {Ivarsen}, {Newton}, {Nysewander}, \& {Reichart}}]{Foley12:11iv}
{Foley} R.~J. {et~al.}, 2012{\natexlab{b}}, \apjl, 753, L5

\bibitem[{{Foley} {et~al}\mbox{.}(2011){Foley}, {Sanders}, \&
  {Kirshner}}]{Foley11:vgrad}
{Foley} R.~J., {Sanders} N.~E., {Kirshner} R.~P., 2011, \apj, 742, 89

\bibitem[{{Foley} {et~al}\mbox{.}(2012{\natexlab{c}}){Foley}, {Simon}, {Burns},
  {Gal-Yam}, {Hamuy}, {Kirshner}, {Morrell}, {Phillips}, {Shields}, \&
  {Sternberg}}]{Foley12:csm}
{Foley} R.~J. {et~al.}, 2012{\natexlab{c}}, \apj, 752, 101

\bibitem[{{Fossey} {et~al}\mbox{.}(2014){Fossey}, {Cooke}, {Pollack}, {Wilde},
  \& {Wright}}]{Fossey14}
{Fossey} J., {Cooke} B., {Pollack} G., {Wilde} M., {Wright} T., 2014, Central
  Bureau Electronic Telegrams, 3792, 1

\bibitem[{{F\"{u}resz}(2008)}]{Furesz08}
{F\"{u}resz} G., 2008, PhD thesis, Univ. of Szeged, Hungary

\bibitem[{{Ganeshalingam} {et~al}\mbox{.}(2010){Ganeshalingam}, {Li},
  {Filippenko}, {Anderson}, {Foster}, {Gates}, {Griffith}, {Grigsby},
  {Joubert}, {Leja}, {Lowe}, {Macomber}, {Pritchard}, {Thrasher}, \&
  {Winslow}}]{Ganeshalingam10}
{Ganeshalingam} M. {et~al.}, 2010, \apjs, 190, 418

\bibitem[{{Garnavich} {et~al}\mbox{.}(2013){Garnavich}, {Milne}, {Bryngelson},
  \& {Leising}}]{Garnavich13}
{Garnavich} P.~M., {Milne} P., {Bryngelson} G.~L., {Leising} M.~D., 2013, in
  American Astronomical Society Meeting Abstracts, Vol. 222, American
  Astronomical Society Meeting Abstracts, p. 209.04

\bibitem[{{Gerke} {et~al}\mbox{.}(2011){Gerke}, {Kochanek}, {Prieto}, {Stanek},
  \& {Macri}}]{Gerke11}
{Gerke} J.~R., {Kochanek} C.~S., {Prieto} J.~L., {Stanek} K.~Z., {Macri} L.~M.,
  2011, \apj, 743, 176

\bibitem[{{Gerke} {et~al}\mbox{.}(2014){Gerke}, {Kochanek}, \&
  {Stanek}}]{Gerke14}
{Gerke} J.~R., {Kochanek} C.~S., {Stanek} K.~Z., 2014, The Astronomer's
  Telegram, 5808, 1

\bibitem[{{Goobar}(2008)}]{Goobar08}
{Goobar} A., 2008, \apjl, 686, L103

\bibitem[{{Goobar} {et~al}\mbox{.}(2014){Goobar}, {Johansson}, {Amanullah},
  {Cao}, {Perley}, {Kasliwal}, {Ferretti}, {Nugent}, {Harris}, {Gal-Yam},
  {Ofek}, {Tendulkar}, {Dennefeld}, {Valenti}, {Arcavi}, {Banerjee},
  {Venkataraman}, {Joshi}, {Ashok}, {Cenko}, {Diaz}, {Fremling}, {Horesh},
  {Howell}, {Kulkarni}, {Papadogiannakis}, {Petrushevska}, {Sand}, {Sollerman},
  {Stanishev}, {Bloom}, {Surace}, {Dupuy}, \& {Liu}}]{Goobar14}
{Goobar} A. {et~al.}, 2014, \apjl, 784, L12

\bibitem[{{Guy} {et~al}\mbox{.}(2005){Guy}, {Astier}, {Nobili}, {Regnault}, \&
  {Pain}}]{Guy05}
{Guy} J., {Astier} P., {Nobili} S., {Regnault} N., {Pain} R., 2005, \aap, 443,
  781

\bibitem[{{Hicken} {et~al}\mbox{.}(2009{\natexlab{a}}){Hicken}, {Challis},
  {Jha}, {Kirshner}, {Matheson}, {Modjaz}, {Rest}, {Michael Wood-Vasey},
  {Bakos}, {Barton}, {Berlind}, {Bragg}, {Brice{\~n}o}, {Brown}, {Caldwell},
  {Calkins}, {Cho}, {Ciupik}, {Contreras}, {Dendy}, {Dosaj}, {Durham},
  {Eriksen}, {Esquerdo}, {Everett}, {Falco}, {Fernandez}, {Gaba}, {Garnavich},
  {Graves}, {Green}, {Groner}, {Hergenrother}, {Holman}, {Hradecky}, {Huchra},
  {Hutchison}, {Jerius}, {Jordan}, {Kilgard}, {Krauss}, {Luhman}, {Macri},
  {Marrone}, {McDowell}, {McIntosh}, {McNamara}, {Megeath}, {Mochejska},
  {Munoz}, {Muzerolle}, {Naranjo}, {Narayan}, {Pahre}, {Peters}, {Peterson},
  {Rines}, {Ripman}, {Roussanova}, {Schild}, {Sicilia-Aguilar}, {Sokoloski},
  {Smalley}, {Smith}, {Spahr}, {Stanek}, {Barmby}, {Blondin}, {Stubbs},
  {Szentgyorgyi}, {Torres}, {Vaz}, {Vikhlinin}, {Wang}, {Westover}, {Woods}, \&
  {Zhao}}]{Hicken09:lc}
{Hicken} M. {et~al.}, 2009{\natexlab{a}}, \apj, 700, 331

\bibitem[{{Hicken} {et~al}\mbox{.}(2009{\natexlab{b}}){Hicken}, {Wood-Vasey},
  {Blondin}, {Challis}, {Jha}, {Kelly}, {Rest}, \& {Kirshner}}]{Hicken09:de}
{Hicken} M., {Wood-Vasey} W.~M., {Blondin} S., {Challis} P., {Jha} S., {Kelly}
  P.~L., {Rest} A., {Kirshner} R.~P., 2009{\natexlab{b}}, \apj, 700, 1097

\bibitem[{{Horesh} {et~al}\mbox{.}(2012){Horesh}, {Kulkarni}, {Fox},
  {Carpenter}, {Kasliwal}, {Ofek}, {Quimby}, {Gal-Yam}, {Cenko}, {de Bruyn},
  {Kamble}, {Wijers}, {van der Horst}, {Kouveliotou}, {Podsiadlowski},
  {Sullivan}, {Maguire}, {Howell}, {Nugent}, {Gehrels}, {Law}, {Poznanski}, \&
  {Shara}}]{Horesh12}
{Horesh} A. {et~al.}, 2012, \apj, 746, 21

\bibitem[{{Hsiao} {et~al}\mbox{.}(2007){Hsiao}, {Conley}, {Howell}, {Sullivan},
  {Pritchet}, {Carlberg}, {Nugent}, \& {Phillips}}]{Hsiao07}
{Hsiao} E.~Y., {Conley} A., {Howell} D.~A., {Sullivan} M., {Pritchet} C.~J.,
  {Carlberg} R.~G., {Nugent} P.~E., {Phillips} M.~M., 2007, \apj, 663, 1187

\bibitem[{{Hutton} {et~al}\mbox{.}(2014){Hutton}, {Ferreras}, {Wu}, {Kuin},
  {Breeveld}, {Yershov}, {Cropper}, \& {Page}}]{Hutton14}
{Hutton} S., {Ferreras} I., {Wu} K., {Kuin} P., {Breeveld} A., {Yershov} V.,
  {Cropper} M., {Page} M., 2014, \mnras, 440, 150

\bibitem[{{Itoh} {et~al}\mbox{.}(2014){Itoh}, {Takaki}, {Ui}, {Kawabata}, \&
  M.}]{Itoh14}
{Itoh} R., {Takaki} K., {Ui} T., {Kawabata} K.~S., M. Y., 2014, Central Bureau
  Electronic Telegrams, 3792, 1

\bibitem[{{Jha} {et~al}\mbox{.}(2007){Jha}, {Riess}, \& {Kirshner}}]{Jha07}
{Jha} S., {Riess} A.~G., {Kirshner} R.~P., 2007, \apj, 659, 122

\bibitem[{{Johansson} {et~al}\mbox{.}(2013){Johansson}, {Amanullah}, \&
  {Goobar}}]{Johansson13}
{Johansson} J., {Amanullah} R., {Goobar} A., 2013, \mnras, 431, L43

\bibitem[{{Kanneganti} {et~al}\mbox{.}(2009){Kanneganti}, {Park}, {Skrutskie},
  {Wilson}, {Nelson}, {Smith}, \& {Lam}}]{Kanneganti09}
{Kanneganti} S., {Park} C., {Skrutskie} M.~F., {Wilson} J.~C., {Nelson} M.~J.,
  {Smith} A.~W., {Lam} C.~R., 2009, \pasp, 121, 885

\bibitem[{{Kattner} {et~al}\mbox{.}(2012){Kattner}, {Leonard}, {Burns},
  {Phillips}, {Folatelli}, {Morrell}, {Stritzinger}, {Hamuy}, {Freedman},
  {Persson}, {Roth}, \& {Suntzeff}}]{Kattner12}
{Kattner} S. {et~al.}, 2012, \pasp, 124, 114

\bibitem[{{Kelly} {et~al}\mbox{.}(2014){Kelly}, {Fox}, {Filippenko}, {Cenko},
  {Prato}, {Schaefer}, {Shen}, {Zheng}, {Graham}, \& {Tucker}}]{Kelly14}
{Kelly} P.~L. {et~al.}, 2014, ArXiv e-prints

\bibitem[{{Krisciunas} {et~al}\mbox{.}(2004){Krisciunas}, {Phillips}, \&
  {Suntzeff}}]{Krisciunas04:hubble}
{Krisciunas} K., {Phillips} M.~M., {Suntzeff} N.~B., 2004, \apjl, 602, L81

\bibitem[{{Krisciunas} {et~al}\mbox{.}(2006){Krisciunas}, {Prieto},
  {Garnavich}, {Riley}, {Rest}, {Stubbs}, \& {McMillan}}]{Krisciunas06}
{Krisciunas} K., {Prieto} J.~L., {Garnavich} P.~M., {Riley} J.-L.~G., {Rest}
  A., {Stubbs} C., {McMillan} R., 2006, \aj, 131, 1639

\bibitem[{{Krist} {et~al}\mbox{.}(2011){Krist}, {Hook}, \& {Stoehr}}]{Krist11}
{Krist} J.~E., {Hook} R.~N., {Stoehr} F., 2011, in Society of Photo-Optical
  Instrumentation Engineers (SPIE) Conference Series, Vol. 8127, Society of
  Photo-Optical Instrumentation Engineers (SPIE) Conference Series

\bibitem[{{Ma} {et~al}\mbox{.}(2014){Ma}, {Wei}, {Shang}, {Wang}, \&
  {Wang}}]{Ma14}
{Ma} B., {Wei} P., {Shang} Z., {Wang} L., {Wang} X., 2014, The Astronomer's
  Telegram, 5794, 1

\bibitem[{{Maguire} {et~al}\mbox{.}(2012){Maguire}, {Sullivan}, {Ellis},
  {Nugent}, {Howell}, {Gal-Yam}, {Cooke}, {Mazzali}, {Pan}, {Dilday}, {Thomas},
  {Arcavi}, {Ben-Ami}, {Bersier}, {Bianco}, {Fulton}, {Hook}, {Horesh},
  {Hsiao}, {James}, {Podsiadlowski}, {Walker}, {Yaron}, {Kasliwal}, {Laher},
  {Law}, {Ofek}, {Poznanski}, \& {Surace}}]{Maguire12}
{Maguire} K. {et~al.}, 2012, \mnras, 426, 2359

\bibitem[{{Maguire} {et~al}\mbox{.}(2013){Maguire}, {Sullivan}, {Patat},
  {Gal-Yam}, {Hook}, {Dhawan}, {Howell}, {Mazzali}, {Nugent}, {Pan},
  {Podsiadlowski}, {Simon}, {Sternberg}, {Valenti}, {Baltay}, {Bersier},
  {Blagorodnova}, {Chen}, {Ellman}, {Feindt}, {F{\"o}rster}, {Fraser},
  {Gonz{\'a}lez-Gait{\'a}n}, {Graham}, {Guti{\'e}rrez}, {Hachinger},
  {Hadjiyska}, {Inserra}, {Knapic}, {Laher}, {Leloudas}, {Margheim},
  {McKinnon}, {Molinaro}, {Morrell}, {Ofek}, {Rabinowitz}, {Rest}, {Sand},
  {Smareglia}, {Smartt}, {Taddia}, {Walker}, {Walton}, \& {Young}}]{Maguire13}
{Maguire} K. {et~al.}, 2013, \mnras, 436, 222

\bibitem[{{Maksym} {et~al}\mbox{.}(2014){Maksym}, {Irwin}, {Keel}, {Burke}, \&
  {Schawinski}}]{Maksym14}
{Maksym} W.~P., {Irwin} J.~A., {Keel} W.~C., {Burke} D., {Schawinski} K., 2014,
  The Astronomer's Telegram, 5798, 1

\bibitem[{{Mandel} {et~al}\mbox{.}(2014){Mandel}, {Foley}, \&
  {Kirshner}}]{Mandel14}
{Mandel} K.~S., {Foley} R.~J., {Kirshner} R.~P., 2014, ArXiv e-prints

\bibitem[{{Mandel} {et~al}\mbox{.}(2011){Mandel}, {Narayan}, \&
  {Kirshner}}]{Mandel11}
{Mandel} K.~S., {Narayan} G., {Kirshner} R.~P., 2011, \apj, 731, 120

\bibitem[{{Mandel} {et~al}\mbox{.}(2009){Mandel}, {Wood-Vasey}, {Friedman}, \&
  {Kirshner}}]{Mandel09}
{Mandel} K.~S., {Wood-Vasey} W.~M., {Friedman} A.~S., {Kirshner} R.~P., 2009,
  \apj, 704, 629

\bibitem[{{Margutti} {et~al}\mbox{.}(2014){Margutti}, {Parrent}, {Kamble},
  {Soderberg}, {Foley}, {Milisavljevic}, {Drout}, \& {Kirshner}}]{Margutti14}
{Margutti} R., {Parrent} J., {Kamble} A., {Soderberg} A.~M., {Foley} R.~J.,
  {Milisavljevic} D., {Drout} M.~R., {Kirshner} R., 2014, ArXiv e-prints

\bibitem[{{Margutti} {et~al}\mbox{.}(2012){Margutti}, {Soderberg}, {Chomiuk},
  {Chevalier}, {Hurley}, {Milisavljevic}, {Foley}, {Hughes}, {Slane},
  {Fransson}, {Moe}, {Barthelmy}, {Boynton}, {Briggs}, {Connaughton}, {Costa},
  {Cummings}, {Del Monte}, {Enos}, {Fellows}, {Feroci}, {Fukazawa}, {Gehrels},
  {Goldsten}, {Golovin}, {Hanabata}, {Harshman}, {Krimm}, {Litvak},
  {Makishima}, {Marisaldi}, {Mitrofanov}, {Murakami}, {Ohno}, {Palmer},
  {Sanin}, {Starr}, {Svinkin}, {Takahashi}, {Tashiro}, {Terada}, \&
  {Yamaoka}}]{Margutti12}
{Margutti} R. {et~al.}, 2012, \apj, 751, 134

\bibitem[{{Marion} {et~al}\mbox{.}(2014){Marion}, {Sand}, {Hsiao}, {Banerjee},
  {Valenti}, {Stritzinger}, {Vink{\'o}}, {Joshi}, {Venkataraman}, {Ashok},
  {Amanullah}, {Binzel}, {Bochanski}, {Bryngelson}, {Burns}, {Drozdov},
  {Fieber-Beyer}, {Graham}, {Howell}, {Johansson}, {Kirshner}, {Milne},
  {Parrent}, {Silverman}, {Vervack}, \& {Wheeler}}]{Marion14}
{Marion} G.~H. {et~al.}, 2014, ArXiv e-prints

\bibitem[{{Matheson} {et~al}\mbox{.}(2012){Matheson}, {Joyce}, {Allen}, {Saha},
  {Silva}, {Wood-Vasey}, {Adams}, {Anderson}, {Beck}, {Bentz}, {Bershady},
  {Binkert}, {Butler}, {Camarata}, {Eigenbrot}, {Everett}, {Gallagher},
  {Garnavich}, {Glikman}, {Harbeck}, {Hargis}, {Herbst}, {Horch}, {Howell},
  {Jha}, {Kaczmarek}, {Knezek}, {Manne-Nicholas}, {Mathieu}, {Meixner},
  {Milliman}, {Power}, {Rajagopal}, {Reetz}, {Rhode}, {Schechtman-Rook},
  {Schwamb}, {Schweiker}, {Simmons}, {Simon}, {Summers}, {Young}, {Weyant},
  {Wilcots}, {Will}, \& {Williams}}]{Matheson12}
{Matheson} T. {et~al.}, 2012, \apj, 754, 19

\bibitem[{{Mazzali} {et~al}\mbox{.}(2014){Mazzali}, {Sullivan}, {Hachinger},
  {Ellis}, {Nugent}, {Howell}, {Gal-Yam}, {Maguire}, {Cooke}, {Thomas},
  {Nomoto}, \& {Walker}}]{Mazzali14}
{Mazzali} P.~A. {et~al.}, 2014, \mnras, 439, 1959

\bibitem[{{Nielsen} {et~al}\mbox{.}(2014){Nielsen}, {Gilfanov}, {Woods}, \&
  {Nelemans}}]{Nielsen14}
{Nielsen} M.~T.~B., {Gilfanov} M., {Woods} T.~E., {Nelemans} G., 2014, The
  Astronomer's Telegram, 5799, 1

\bibitem[{{Nobili} {et~al}\mbox{.}(2005){Nobili}, {Amanullah}, {Garavini},
  {Goobar}, {Lidman}, {Stanishev}, {Aldering}, {Antilogus}, {Astier}, {Burns},
  {Conley}, {Deustua}, {Ellis}, {Fabbro}, {Fadeyev}, {Folatelli}, {Gibbons},
  {Goldhaber}, {Groom}, {Hook}, {Howell}, {Kim}, {Knop}, {Nugent}, {Pain},
  {Perlmutter}, {Quimby}, {Raux}, {Regnault}, {Ruiz-Lapuente}, {Sainton},
  {Schahmaneche}, {Smith}, {Spadafora}, {Thomas}, {Wang}, \& {Supernova
  Cosmology Project}}]{Nobili05}
{Nobili} S. {et~al.}, 2005, \aap, 437, 789

\bibitem[{{Nugent} {et~al}\mbox{.}(2002){Nugent}, {Kim}, \&
  {Perlmutter}}]{Nugent02}
{Nugent} P., {Kim} A., {Perlmutter} S., 2002, \pasp, 114, 803

\bibitem[{{Nugent} {et~al}\mbox{.}(2011){Nugent}, {Sullivan}, {Cenko},
  {Thomas}, {Kasen}, {Howell}, {Bersier}, {Bloom}, {Kulkarni}, {Kandrashoff},
  {Filippenko}, {Silverman}, {Marcy}, {Howard}, {Isaacson}, {Maguire},
  {Suzuki}, {Tarlton}, {Pan}, {Bildsten}, {Fulton}, {Parrent}, {Sand},
  {Podsiadlowski}, {Bianco}, {Dilday}, {Graham}, {Lyman}, {James}, {Kasliwal},
  {Law}, {Quimby}, {Hook}, {Walker}, {Mazzali}, {Pian}, {Ofek}, {Gal-Yam}, \&
  {Poznanski}}]{Nugent11}
{Nugent} P.~E. {et~al.}, 2011, \nat, 480, 344

\bibitem[{{O'Donnell}(1994)}]{Odonnell94}
{O'Donnell} J.~E., 1994, \apj, 422, 158

\bibitem[{{Patat} {et~al}\mbox{.}(2006){Patat}, {Benetti}, {Cappellaro}, \&
  {Turatto}}]{Patat06}
{Patat} F., {Benetti} S., {Cappellaro} E., {Turatto} M., 2006, \mnras, 369,
  1949

\bibitem[{{Patat} {et~al}\mbox{.}(2007){Patat}, {Chandra}, {Chevalier},
  {Justham}, {Podsiadlowski}, {Wolf}, {Gal-Yam}, {Pasquini}, {Crawford},
  {Mazzali}, {Pauldrach}, {Nomoto}, {Benetti}, {Cappellaro}, {Elias-Rosa},
  {Hillebrandt}, {Leonard}, {Pastorello}, {Renzini}, {Sabbadin}, {Simon}, \&
  {Turatto}}]{Patat07:06x}
{Patat} F. {et~al.}, 2007, Science, 317, 924

\bibitem[{{Patat} {et~al}\mbox{.}(2011){Patat}, {Chugai}, {Podsiadlowski},
  {Mason}, {Melo}, \& {Pasquini}}]{Patat11}
{Patat} F., {Chugai} N.~N., {Podsiadlowski} P., {Mason} E., {Melo} C.,
  {Pasquini} L., 2011, \aap, 530, A63

\bibitem[{{Patat} {et~al}\mbox{.}(2013){Patat}, {Cordiner}, {Cox}, {Anderson},
  {Harutyunyan}, {Kotak}, {Palaversa}, {Stanishev}, {Tomasella}, {Benetti},
  {Goobar}, {Pastorello}, \& {Sollerman}}]{Patat13}
{Patat} F. {et~al.}, 2013, \aap, 549, A62

\bibitem[{{Patat} {et~al}\mbox{.}(2014){Patat}, {Taubenberger}, {Baade},
  {Hoeflich}, {Maund}, {Reilly}, {Spyromilio}, {Wang}, {Wheeler}, \&
  {Zelaya}}]{Patat14}
{Patat} F. {et~al.}, 2014, The Astronomer's Telegram, 5830, 1

\bibitem[{{Perlmutter} {et~al}\mbox{.}(1999){Perlmutter}, {Aldering},
  {Goldhaber}, {Knop}, {Nugent}, {Castro}, {Deustua}, {Fabbro}, {Goobar},
  {Groom}, {Hook}, {Kim}, {Kim}, {Lee}, {Nunes}, {Pain}, {Pennypacker},
  {Quimby}, {Lidman}, {Ellis}, {Irwin}, {McMahon}, {Ruiz-Lapuente}, {Walton},
  {Schaefer}, {Boyle}, {Filippenko}, {Matheson}, {Fruchter}, {Panagia},
  {Newberg}, \& {Couch}}]{Perlmutter99}
{Perlmutter} S. {et~al.}, 1999, \apj, 517, 565

\bibitem[{{Phillips} {et~al}\mbox{.}(1999){Phillips}, {Lira}, {Suntzeff},
  {Schommer}, {Hamuy}, \& {Maza}}]{Phillips99}
{Phillips} M.~M., {Lira} P., {Suntzeff} N.~B., {Schommer} R.~A., {Hamuy} M.,
  {Maza} J., 1999, \aj, 118, 1766

\bibitem[{{Phillips} {et~al}\mbox{.}(2013){Phillips}, {Simon}, {Morrell},
  {Burns}, {Cox}, {Foley}, {Karakas}, {Patat}, {Sternberg}, {Williams},
  {Gal-Yam}, {Hsiao}, {Leonard}, {Persson}, {Stritzinger}, {Thompson},
  {Campillay}, {Contreras}, {Folatelli}, {Freedman}, {Hamuy}, {Roth},
  {Shields}, {Suntzeff}, {Chomiuk}, {Ivans}, {Madore}, {Penprase}, {Perley},
  {Pignata}, {Preston}, \& {Soderberg}}]{Phillips13}
{Phillips} M.~M. {et~al.}, 2013, \apj, 779, 38

\bibitem[{{Poznanski} {et~al}\mbox{.}(2011){Poznanski}, {Ganeshalingam},
  {Silverman}, \& {Filippenko}}]{Poznanski11}
{Poznanski} D., {Ganeshalingam} M., {Silverman} J.~M., {Filippenko} A.~V.,
  2011, \mnras, 415, L81

\bibitem[{{Prieto} {et~al}\mbox{.}(2006){Prieto}, {Rest}, \&
  {Suntzeff}}]{Prieto06}
{Prieto} J.~L., {Rest} A., {Suntzeff} N.~B., 2006, \apj, 647, 501

\bibitem[{{Rayner} {et~al}\mbox{.}(2003){Rayner}, {Toomey}, {Onaka}, {Denault},
  {Stahlberger}, {Vacca}, {Cushing}, \& {Wang}}]{Rayner03}
{Rayner} J.~T., {Toomey} D.~W., {Onaka} P.~M., {Denault} A.~J., {Stahlberger}
  W.~E., {Vacca} W.~D., {Cushing} M.~C., {Wang} S., 2003, \pasp, 115, 362

\bibitem[{{Rest} {et~al}\mbox{.}(2013){Rest}, {Scolnic}, {Foley}, {Huber},
  {Chornock}, {Narayan}, {Tonry}, {Berger}, {Soderberg}, {Stubbs}, {Riess},
  {Kirshner}, {Smartt}, {Schlafly}, {Rodney}, {Botticella}, {Brout}, {Challis},
  {Czekala}, {Drout}, {Hudson}, {Kotak}, {Leibler}, {Lunnan}, {Marion},
  {McCrum}, {Milisavljevic}, {Pastorello}, {Sanders}, {Smith}, {Stafford},
  {Thilker}, {Valenti}, {Wood-Vasey}, {Zheng}, {Burgett}, {Chambers},
  {Denneau}, {Draper}, {Flewelling}, {Hodapp}, {Kaiser}, {Kudritzki},
  {Magnier}, {Metcalfe}, {Price}, {Sweeney}, {Wainscoat}, \& {Waters}}]{Rest13}
{Rest} A. {et~al.}, 2013, ArXiv e-prints

\bibitem[{{Richmond} \& {Smith}(2012)}]{Richmond12}
{Richmond} M.~W., {Smith} H.~A., 2012, Journal of the American Association of
  Variable Star Observers (JAAVSO), 40, 872

\bibitem[{{Riess} {et~al}\mbox{.}(1998){Riess}, {Filippenko}, {Challis},
  {Clocchiatti}, {Diercks}, {Garnavich}, {Gilliland}, {Hogan}, {Jha},
  {Kirshner}, {Leibundgut}, {Phillips}, {Reiss}, {Schmidt}, {Schommer},
  {Smith}, {Spyromilio}, {Stubbs}, {Suntzeff}, \& {Tonry}}]{Riess98:Lambda}
{Riess} A.~G. {et~al.}, 1998, \aj, 116, 1009

\bibitem[{{Riess} {et~al}\mbox{.}(1996){Riess}, {Press}, \&
  {Kirshner}}]{Riess96}
{Riess} A.~G., {Press} W.~H., {Kirshner} R.~P., 1996, \apj, 473, 88

\bibitem[{{Schlafly} \& {Finkbeiner}(2011)}]{Schlafly11}
{Schlafly} E.~F., {Finkbeiner} D.~P., 2011, \apj, 737, 103

\bibitem[{{Schlegel} {et~al}\mbox{.}(1998){Schlegel}, {Finkbeiner}, \&
  {Davis}}]{Schlegel98}
{Schlegel} D.~J., {Finkbeiner} D.~P., {Davis} M., 1998, \apj, 500, 525

\bibitem[{{Scolnic} {et~al}\mbox{.}(2013){Scolnic}, {Rest}, {Riess}, {Huber},
  {Foley}, {Brout}, {Chornock}, {Narayan}, {Tonry}, {Berger}, {Soderberg},
  {Stubbs}, {Kirshner}, {Rodney}, {Smartt}, {Schlafly}, {Botticella},
  {Challis}, {Czekal}, {Drout}, {Hudson}, {Kotak}, {Leibler}, {Lunnan},
  {Marion}, {McCrum}, {Milisavljevic}, {Pastorello}, {Sanders}, {Smith},
  {Stafford}, {Thilker}, {Valenti}, {Wood-Vasey}, {Zheng}, {Burgett},
  {Chambers}, {Denneau}, {Draper}, {Flewelling}, {Hodapp}, {Kaiser},
  {Kudritzki}, {Magnier}, {Metcalfe}, {Price}, {Sweeney}, {Wainscoat}, \&
  {Waters}}]{Scolnic14:ps1}
{Scolnic} D. {et~al.}, 2013, ArXiv e-prints

\bibitem[{{Scolnic} {et~al}\mbox{.}(2014){Scolnic}, {Riess}, {Foley}, {Rest},
  {Rodney}, {Brout}, \& {Jones}}]{Scolnic14:col}
{Scolnic} D.~M., {Riess} A.~G., {Foley} R.~J., {Rest} A., {Rodney} S.~A.,
  {Brout} D.~J., {Jones} D.~O., 2014, \apj, 780, 37

\bibitem[{{Simon} {et~al}\mbox{.}(2009){Simon}, {Gal-Yam}, {Gnat}, {Quimby},
  {Ganeshalingam}, {Silverman}, {Blondin}, {Li}, {Filippenko}, {Wheeler},
  {Kirshner}, {Patat}, {Nugent}, {Foley}, {Vogt}, {Butler}, {Peek},
  {Rosolowsky}, {Herczeg}, {Sauer}, \& {Mazzali}}]{Simon09}
{Simon} J.~D. {et~al.}, 2009, \apj, 702, 1157

\bibitem[{{Skrutskie} {et~al}\mbox{.}(2006){Skrutskie}, {Cutri}, {Stiening},
  {Weinberg}, {Schneider}, {Carpenter}, {Beichman}, {Capps}, {Chester},
  {Elias}, {Huchra}, {Liebert}, {Lonsdale}, {Monet}, {Price}, {Seitzer},
  {Jarrett}, {Kirkpatrick}, {Gizis}, {Howard}, {Evans}, {Fowler}, {Fullmer},
  {Hurt}, {Light}, {Kopan}, {Marsh}, {McCallon}, {Tam}, {Van Dyk}, \&
  {Wheelock}}]{Skrutskie06}
{Skrutskie} M.~F. {et~al.}, 2006, \aj, 131, 1163

\bibitem[{{Sternberg} {et~al}\mbox{.}(2011){Sternberg}, {Gal-Yam}, {Simon},
  {Leonard}, {Quimby}, {Phillips}, {Morrell}, {Thompson}, {Ivans}, {Marshall},
  {Filippenko}, {Marcy}, {Bloom}, {Patat}, {Foley}, {Yong}, {Penprase},
  {Beeler}, {Allende Prieto}, \& {Stringfellow}}]{Sternberg11}
{Sternberg} A. {et~al.}, 2011, Science, 333, 856

\bibitem[{{Stetson}(1987)}]{Stetson87}
{Stetson} P.~B., 1987, \pasp, 99, 191

\bibitem[{{Stritzinger} {et~al}\mbox{.}(2002){Stritzinger}, {Hamuy},
  {Suntzeff}, {Smith}, {Phillips}, {Maza}, {Strolger}, {Antezana},
  {Gonz{\'a}lez}, {Wischnjewsky}, {Candia}, {Espinoza}, {Gonz{\'a}lez},
  {Stubbs}, {Becker}, {Rubenstein}, \& {Galaz}}]{Stritzinger02}
{Stritzinger} M. {et~al.}, 2002, \aj, 124, 2100

\bibitem[{{Tsvetkov} {et~al}\mbox{.}(2014){Tsvetkov}, {Metlov}, {Shugarov},
  {Tarasova}, \& {Pavlyuk}}]{Tsvetkov14}
{Tsvetkov} D.~Y., {Metlov} V.~G., {Shugarov} S.~Y., {Tarasova} T.~N., {Pavlyuk}
  N.~N., 2014, ArXiv e-prints

\bibitem[{{Tully}(1988)}]{Tully88}
{Tully} R.~B., 1988, {Nearby galaxies catalog}

\bibitem[{{Vacca} {et~al}\mbox{.}(2003){Vacca}, {Cushing}, \&
  {Rayner}}]{Vacca03}
{Vacca} W.~D., {Cushing} M.~C., {Rayner} J.~T., 2003, \pasp, 115, 389

\bibitem[{{Wang}(2005)}]{Wang05}
{Wang} L., 2005, \apjl, 635, L33

\bibitem[{{Wang} {et~al}\mbox{.}(2009){Wang}, {Filippenko}, {Ganeshalingam},
  {Li}, {Silverman}, {Wang}, {Chornock}, {Foley}, {Gates}, {Macomber},
  {Serduke}, {Steele}, \& {Wong}}]{Wang09:2pop}
{Wang} X. {et~al.}, 2009, \apjl, 699, L139

\bibitem[{{Wang} {et~al}\mbox{.}(2008){Wang}, {Li}, {Filippenko}, {Krisciunas},
  {Suntzeff}, {Li}, {Zhang}, {Deng}, {Foley}, {Ganeshalingam}, {Li}, {Lou},
  {Qiu}, {Shang}, {Silverman}, {Zhang}, \& {Zhang}}]{Wang08:06x}
{Wang} X. {et~al.}, 2008, \apj, 675, 626

\bibitem[{{Welty} {et~al}\mbox{.}(2014){Welty}, {Ritchey}, {Dahlstrom}, \&
  {York}}]{Welty14}
{Welty} D.~E., {Ritchey} A.~M., {Dahlstrom} J.~A., {York} D.~G., 2014, ArXiv
  e-prints

\bibitem[{{Wood-Vasey} {et~al}\mbox{.}(2008){Wood-Vasey}, {Friedman}, {Bloom},
  {Hicken}, {Modjaz}, {Kirshner}, {Starr}, {Blake}, {Falco}, {Szentgyorgyi},
  {Challis}, {Blondin}, {Mandel}, \& {Rest}}]{Wood-Vasey08}
{Wood-Vasey} W.~M. {et~al.}, 2008, \apj, 689, 377

\bibitem[{{Yun} {et~al}\mbox{.}(1994){Yun}, {Ho}, \& {Lo}}]{Yun94}
{Yun} M.~S., {Ho} P.~T.~P., {Lo} K.~Y., 1994, \nat, 372, 530

\bibitem[{{Zheng} {et~al}\mbox{.}(2014){Zheng}, {Shivvers}, {Filippenko},
  {Itagaki}, {Clubb}, {Fox}, {Graham}, {Kelly}, \& {Mauerhan}}]{Zheng14}
{Zheng} W. {et~al.}, 2014, \apjl, 783, L24

\bibitem[{{Zheng} {et~al}\mbox{.}(2013){Zheng}, {Silverman}, {Filippenko},
  {Kasen}, {Nugent}, {Graham}, {Wang}, {Valenti}, {Ciabattari}, {Kelly}, {Fox},
  {Shivvers}, {Clubb}, {Cenko}, {Balam}, {Howell}, {Hsiao}, {Li}, {Marion},
  {Sand}, {Vinko}, {Wheeler}, \& {Zhang}}]{Zheng13}
{Zheng} W. {et~al.}, 2013, \apjl, 778, L15

\end{thebibliography}

\begin{deluxetable}{lcr}
\tabletypesize{\footnotesize}
\tablewidth{0pt}
\tablecaption{{\it HST} Photometry\label{t:hstphot}}
\tablehead{
\colhead{MJD} &
\colhead{Filter} &
\colhead{Magnitude}}

\startdata

56685.119 & F218W & 18.158 (021) \\
56688.845 & F218W & 18.020 (019) \\
56692.160 & F218W & 18.082 (017) \\
56696.859 & F218W & 18.309 (023) \\
56702.568 & F218W & 18.861 (019) \\
56712.989 & F218W & 19.767 (038) \\
56723.031 & F218W & 20.505 (043) \\
56685.122 & F225W & 18.637 (019) \\
56688.847 & F225W & 18.498 (017) \\
56692.162 & F225W & 18.619 (019) \\
56696.862 & F225W & 18.946 (020) \\
56702.571 & F225W & 19.304 (021) \\
56712.991 & F225W & 19.541 (025) \\
56723.034 & F225W & 19.737 (017) \\
56685.124 & F275W & 16.550 (015) \\
56688.849 & F275W & 16.440 (014) \\
56692.164 & F275W & 16.743 (017) \\
56696.864 & F275W & 17.291 (018) \\
56702.573 & F275W & 18.086 (021) \\
56712.993 & F275W & 19.188 (031) \\
56723.037 & F275W & 19.747 (020) \\
56685.125 & F336W & 13.096 (008) \\
56688.850 & F336W & 13.037 (008) \\
56692.166 & F336W & 13.359 (009) \\
56696.865 & F336W & 13.843 (012) \\
56702.574 & F336W & 14.702 (006) \\
56712.995 & F336W & 16.048 (010) \\
56723.039 & F336W & 16.824 (010) \\
56688.892 & F438W & 11.876 (004) \\
56696.888 & F438W & 12.228 (003) \\
56702.596 & F438W & 12.629 (006) \\
56713.017 & F438W & 13.807 (010) \\
56685.147 & F467M & 11.808 (007) \\
56692.225 & F467M & 11.649 (004) \\
56688.893 & F555W & 10.817 (002) \\
56696.889 & F555W & 11.069 (001) \\
56702.598 & F555W & 11.278 (002) \\
56713.018 & F555W & 11.778 (003) \\
56685.149 & F631N & 10.036 (008) \\
56692.227 & F631N &  9.940 (005) \\
56688.894 & F814W &  9.799 (002) \\
56696.891 & F814W & 10.131 (001) \\
56702.599 & F814W & 10.357 (002) \\
56713.019 & F814W & 10.196 (002) \\
56685.150 & F845M &  9.639 (003) \\
56692.228 & F845M &  9.664 (002)

\enddata

\end{deluxetable}

\begin{deluxetable}{lcr}
\tabletypesize{\footnotesize}
\tablewidth{0pt}
\tablecaption{KAIT Photometry\label{t:kait}}
\tablehead{
\colhead{MJD} &
\colhead{Filter} &
\colhead{Magnitude}}

\startdata

56679.369 & $B$ & 12.966 (022) \\
56680.366 & $B$ & 12.695 (012) \\
56682.328 & $B$ & 12.329 (009) \\
56683.374 & $B$ & 12.209 (008) \\
56684.357 & $B$ & 12.105 (010) \\
56689.393 & $B$ & 11.888 (011) \\
56690.301 & $B$ & 11.860 (008) \\
56700.431 & $B$ & 12.408 (019) \\
56701.304 & $B$ & 12.426 (018) \\
56702.293 & $B$ & 12.513 (019) \\
56703.344 & $B$ & 12.625 (013) \\
56705.331 & $B$ & 12.850 (010) \\
56706.322 & $B$ & 12.932 (009) \\
56708.267 & $B$ & 13.105 (011) \\
56709.340 & $B$ & 13.208 (009) \\
56710.291 & $B$ & 13.280 (022) \\
56711.248 & $B$ & 13.347 (014) \\
56712.256 & $B$ & 13.442 (016) \\
56713.295 & $B$ & 13.539 (017) \\
56714.260 & $B$ & 13.607 (036) \\
56719.365 & $B$ & 14.073 (014) \\
56721.273 & $B$ & 14.181 (015) \\
56723.249 & $B$ & 14.296 (011) \\
56724.248 & $B$ & 14.335 (013) \\
56725.249 & $B$ & 14.361 (025) \\
56727.252 & $B$ & 14.341 (027) \\
56728.222 & $B$ & 14.425 (015) \\
56729.271 & $B$ & 14.527 (014) \\
56730.236 & $B$ & 14.549 (016) \\
56731.272 & $B$ & 14.596 (017) \\
56733.285 & $B$ & 14.603 (018) \\
56735.279 & $B$ & 14.680 (016) \\
56737.249 & $B$ & 14.722 (045) \\
56739.286 & $B$ & 14.727 (013) \\
56740.261 & $B$ & 14.716 (020) \\
56741.243 & $B$ & 14.768 (015) \\
56744.316 & $B$ & 14.779 (036) \\
56753.275 & $B$ & 14.976 (017) \\
56755.258 & $B$ & 14.942 (019) \\
56757.254 & $B$ & 14.973 (018) \\
56759.269 & $B$ & 15.049 (143) \\
56761.290 & $B$ & 15.044 (018) \\
56763.278 & $B$ & 15.060 (022) \\
56765.254 & $B$ & 15.074 (018) \\
56767.262 & $B$ & 15.055 (020) \\
56679.369 & $V$ & 11.721 (018) \\
56680.367 & $V$ & 11.482 (013) \\
56682.329 & $V$ & 11.129 (013) \\
56683.375 & $V$ & 11.000 (012) \\
56684.358 & $V$ & 10.892 (012) \\
56689.394 & $V$ & 10.621 (017) \\
56690.301 & $V$ & 10.601 (012) \\
56700.428 & $V$ & 10.975 (037) \\
56701.304 & $V$ & 10.948 (018) \\
56702.293 & $V$ & 10.996 (024) \\
56703.345 & $V$ & 11.104 (013) \\
56705.332 & $V$ & 11.228 (014) \\
56706.323 & $V$ & 11.280 (012) \\
56708.266 & $V$ & 11.330 (012) \\
56709.341 & $V$ & 11.390 (012) \\
56710.292 & $V$ & 11.412 (026) \\
56711.248 & $V$ & 11.430 (016) \\
56712.257 & $V$ & 11.469 (025) \\
56713.296 & $V$ & 11.517 (022) \\
56714.261 & $V$ & 11.565 (021) \\
56719.366 & $V$ & 11.848 (014) \\
56721.270 & $V$ & 11.954 (017) \\
56723.250 & $V$ & 12.064 (009) \\
56724.248 & $V$ & 12.122 (011) \\
56725.250 & $V$ & 12.146 (018) \\
56726.298 & $V$ & 12.189 (139) \\
56727.251 & $V$ & 12.227 (008) \\
56728.223 & $V$ & 12.296 (008) \\
56729.271 & $V$ & 12.364 (012) \\
56730.236 & $V$ & 12.393 (011) \\
56731.273 & $V$ & 12.457 (016) \\
56733.285 & $V$ & 12.490 (013) \\
56735.280 & $V$ & 12.580 (014) \\
56737.250 & $V$ & 12.671 (024) \\
56739.287 & $V$ & 12.706 (011) \\
56740.260 & $V$ & 12.688 (011) \\
56741.244 & $V$ & 12.752 (011) \\
56744.317 & $V$ & 12.820 (018) \\
56753.276 & $V$ & 13.115 (018) \\
56755.258 & $V$ & 13.140 (010) \\
56757.254 & $V$ & 13.184 (017) \\
56759.268 & $V$ & 13.252 (042) \\
56761.291 & $V$ & 13.308 (011) \\
56763.279 & $V$ & 13.356 (014) \\
56765.255 & $V$ & 13.377 (009) \\
56767.262 & $V$ & 13.373 (013) \\
56679.370 & $R$ & 11.105 (021) \\
56680.367 & $R$ & 10.859 (013) \\
56682.329 & $R$ & 10.527 (016) \\
56683.375 & $R$ & 10.412 (020) \\
56684.358 & $R$ & 10.331 (017) \\
56689.394 & $R$ & 10.141 (018) \\
56690.302 & $R$ & 10.124 (012) \\
56700.418 & $R$ & 10.536 (025) \\
56701.306 & $R$ & 10.591 (024) \\
56702.294 & $R$ & 10.652 (030) \\
56703.346 & $R$ & 10.735 (017) \\
56705.332 & $R$ & 10.816 (017) \\
56706.323 & $R$ & 10.846 (017) \\
56708.267 & $R$ & 10.827 (015) \\
56709.341 & $R$ & 10.863 (014) \\
56710.292 & $R$ & 10.839 (029) \\
56711.249 & $R$ & 10.832 (021) \\
56712.258 & $R$ & 10.841 (031) \\
56713.296 & $R$ & 10.871 (031) \\
56714.261 & $R$ & 10.902 (022) \\
56719.364 & $R$ & 11.064 (013) \\
56721.271 & $R$ & 11.158 (020) \\
56723.251 & $R$ & 11.269 (009) \\
56724.249 & $R$ & 11.336 (015) \\
56725.250 & $R$ & 11.369 (021) \\
56726.298 & $R$ & 11.374 (094) \\
56727.252 & $R$ & 11.456 (008) \\
56728.224 & $R$ & 11.562 (007) \\
56729.272 & $R$ & 11.614 (010) \\
56730.237 & $R$ & 11.662 (012) \\
56731.274 & $R$ & 11.739 (021) \\
56733.286 & $R$ & 11.779 (016) \\
56735.281 & $R$ & 11.874 (017) \\
56737.250 & $R$ & 11.984 (026) \\
56739.287 & $R$ & 12.030 (015) \\
56740.260 & $R$ & 12.040 (012) \\
56741.244 & $R$ & 12.103 (018) \\
56744.318 & $R$ & 12.178 (024) \\
56753.277 & $R$ & 12.505 (014) \\
56755.259 & $R$ & 12.542 (014) \\
56757.255 & $R$ & 12.608 (019) \\
56759.269 & $R$ & 12.679 (036) \\
56761.291 & $R$ & 12.737 (012) \\
56763.280 & $R$ & 12.807 (024) \\
56765.256 & $R$ & 12.819 (010) \\
56767.263 & $R$ & 12.824 (013) \\
56679.370 & $I$ & 10.626 (026) \\
56680.367 & $I$ & 10.358 (015) \\
56682.330 & $I$ & 10.041 (021) \\
56683.376 & $I$ &  9.911 (024) \\
56684.359 & $I$ &  9.833 (021) \\
56689.395 & $I$ &  9.768 (025) \\
56690.302 & $I$ &  9.787 (014) \\
56700.418 & $I$ & 10.302 (042) \\
56701.302 & $I$ & 10.298 (030) \\
56702.296 & $I$ & 10.337 (037) \\
56703.346 & $I$ & 10.380 (021) \\
56705.333 & $I$ & 10.388 (018) \\
56706.324 & $I$ & 10.390 (025) \\
56708.267 & $I$ & 10.307 (015) \\
56709.341 & $I$ & 10.291 (016) \\
56710.293 & $I$ & 10.231 (034) \\
56711.249 & $I$ & 10.223 (022) \\
56712.258 & $I$ & 10.204 (033) \\
56713.297 & $I$ & 10.188 (034) \\
56714.262 & $I$ & 10.173 (026) \\
56719.365 & $I$ & 10.160 (019) \\
56721.271 & $I$ & 10.210 (030) \\
56723.251 & $I$ & 10.298 (013) \\
56724.249 & $I$ & 10.369 (018) \\
56725.251 & $I$ & 10.443 (025) \\
56726.305 & $I$ & 10.518 (079) \\
56727.252 & $I$ & 10.586 (009) \\
56728.224 & $I$ & 10.651 (007) \\
56729.272 & $I$ & 10.720 (012) \\
56730.237 & $I$ & 10.780 (013) \\
56731.274 & $I$ & 10.851 (025) \\
56733.286 & $I$ & 10.925 (019) \\
56735.281 & $I$ & 11.072 (020) \\
56737.251 & $I$ & 11.174 (032) \\
56739.288 & $I$ & 11.277 (022) \\
56740.260 & $I$ & 11.273 (013) \\
56741.245 & $I$ & 11.369 (022) \\
56744.318 & $I$ & 11.484 (026) \\
56753.277 & $I$ & 11.921 (020) \\
56755.259 & $I$ & 11.990 (018) \\
56757.255 & $I$ & 12.068 (023) \\
56759.271 & $I$ & 12.164 (056) \\
56761.292 & $I$ & 12.224 (015) \\
56763.280 & $I$ & 12.333 (024) \\
56765.256 & $I$ & 12.330 (015) \\
56767.263 & $I$ & 12.315 (022)

\enddata

\end{deluxetable}

\begin{deluxetable}{lrrr}
\tablewidth{0pt}
\tabletypesize{\footnotesize}
\tablecaption{FanCam Photometry \label{t:fancam}}
\tablecolumns{4}
\tablehead{
\colhead{UT} & \colhead{$J$} & \colhead{$H$} & \colhead{$K_{s}$}\\
\colhead{(days)} & \multicolumn{3}{c}{Magnitude}
}
\startdata

20140130.178 &  9.36 (02) &  9.44 (02) &  9.20 (01)\\
20140201.101 &  9.51 (02) &  9.59 (02) &  9.38 (02)\\
20140204.002 &  9.74 (01) &  9.67 (02) &  9.43 (02)\\
20140208.096 & 10.03 (02) &  9.71 (02) &  9.47 (02)\\
20140212.154 & 10.71 (01) &  9.77 (02) &  9.62 (02)\\
20140220.101 & 10.89 (02) &  9.56 (02) &  9.48 (02)\\
20140223.007 & 10.80 (02) &  9.49 (02) &  9.43 (02)\\
20140225.004 & 10.68 (02) &  9.47 (02) &  9.35 (02)\\
20140228.185 & 10.57 (02) &  9.57 (02) &  9.39 (02)\\
20140401.600 & 11.80 (05) & 10.65 (07) & 10.74 (10)\\
20140417.600 & 12.28 (05) & 11.02 (10) & 10.93 (10)

\enddata

\end{deluxetable}

%
%
%

\begin{deluxetable}{rlr}
\tabletypesize{\footnotesize}
\tablewidth{0pt}
\tablecaption{Log of {\it HST}/STIS Spectral Observations\label{t:hstspec}}
\tablehead{
\colhead{Phase\tablenotemark{a}} &
\colhead{UT Date} &
\colhead{Exposure\tablenotemark{b} (s)}}

\startdata

$-6.4$ & 2014 Jan.\ 26.600 &  9573+300+200 \\
$-4.6$ & 2014 Jan.\ 28.443 &  4002+160+100 \\
$-2.5$ & 2014 Jan.\ 30.485 &  4002+160+100 \\
$-0.4$ & 2014 Feb.\  1.610 &  4002+160+100 \\
 +2.7  & 2014 Feb.\  4.731 &  2255+160+100 \\
 +6.5  & 2014 Feb.\  8.525 &  4002+160+100 \\
 +8.4  & 2014 Feb.\ 10.439 &  4002+160+100 \\
+11.3  & 2014 Feb.\ 13.293 &  4002+160+100 \\
+14.4  & 2014 Feb.\ 16.412 &  4002+160+100 \\
+24.1  & 2014 Feb.\ 26.069 & 11828+300+200

\enddata

\tablenotetext{a}{Days since $B$ maximum, 2014 Feb.\ 2.0 (JD
  2,456,690.5).}

\tablenotetext{b}{First, second, and third numbers correspond to the
  time for the G230L, G430L, and G750L gratings, respectively.}

\end{deluxetable}

\onecolumn

\begin{deluxetable}{clrccc}
\tabletypesize{\footnotesize}
\tablewidth{0pt}
\tablecaption{Log of IRTF/SpeX Spectral Observations\label{t:nirspec}}
\tablehead{
\colhead{Phase\tablenotemark{a}} &
\colhead{UT Date} &
\colhead{Exposure\tablenotemark{b} (s)} &
\colhead{SN~2014J} &
\colhead{Standard} &
\colhead{Standard} \\
\colhead{} &
\colhead{} &
\colhead{} &
\colhead{Airmass} &
\colhead{} &
\colhead{Airmass} }

\startdata

$-7.5$ & 2014 Jan.\ 25.45 &  720 & 1.58 & HIP52478 & 1.31 \\
$-6.5$ & 2014 Jan.\ 26.46 &  720 & 1.52 & HIP52478 & 1.30 \\
$-5.6$ & 2014 Jan.\ 27.42 &  960 & 1.64 & HIP52478 & 1.43 \\
$-4.7$ & 2014 Jan.\ 28.30 & 2000 & 2.45 & HIP45590 & 2.25 \\
$-0.6$ & 2014 Feb.\ 01.39 &  840 & 1.67 & HIP52478 & 1.45

\enddata

\tablenotetext{a}{Days since $B$ maximum, 2014 Feb.\ 2.0 (JD
  2,456,690.5).}

\end{deluxetable}

\begin{deluxetable}{rlr}
\tabletypesize{\footnotesize}
\tablewidth{0pt}
\tablecaption{Log of TRES Spectral Observations\label{t:tres}}
\tablehead{
\colhead{Phase\tablenotemark{a}} &
\colhead{UT Date} &
\colhead{Exposure\tablenotemark{b} (s)}}

\startdata

$-10.6$ & 2014 Jan.\ 23.399 & 1800 \\
 $-7.5$ & 2014 Jan.\ 26.463 & 1200 \\
  +7.3  & 2014 Feb.\ 10.323 & 1980 \\
  +9.3  & 2014 Feb.\ 12.315 & 1800 \\
 +11.3  & 2014 Feb.\ 14.329 & 1800 \\
 +12.3  & 2014 Feb.\ 15.329 & 1800 \\
 +15.4  & 2014 Feb.\ 18.413 & 1800 \\
 +17.4  & 2014 Feb.\ 20.433 & 1920 \\
 +41.2  & 2014 Mar.\ 16.243 & 1800 \\
 +47.3  & 2014 Mar.\ 22.302 & 1000

\enddata

\tablenotetext{a}{Days since $B$ maximum, 2014 Feb.\ 2.0 (JD
  2,456,690.5).}

\end{deluxetable}

\begin{deluxetable}{crrcc}
\tabletypesize{\footnotesize}
\tablewidth{0pt}
\tablecaption{Photometric Parameters\label{t:params}}
\tablehead{
\colhead{Filter} &
\colhead{$t_{\rm max}$} &
\colhead{Maximum Brightness} &
\colhead{$\Delta m_{15}$} &
\colhead{$A_{X}$\tablenotemark{a}} \\
\colhead{} &
\colhead{(MJD)} &
\colhead{(mag)} &
\colhead{(mag)} &
\colhead{(mag)}}

\startdata
F218W   & 56689.3 (1)       & 18.02 (03)     & 0.99 (05) & \nodata \\
F225W   & 56688.4 (1)       & 18.50 (03)     & 0.85 (04) & \nodata \\
F275W   & 56687.6 (2)       & 16.42 (03)     & 1.65 (05) & \nodata \\
F336W   & 56687.2 (2)       & 13.02 (03)     & 1.62 (02) & \nodata \\
$B$     & 56690.0 (2)       & 11.85 (02)     & 0.95 (01) & 3.38 (20) \\
$V$     & 56691.0 (6)       & 10.61 (05)     & 0.60 (01) & 2.07 (18) \\
$R$     & 56690.9 (8)       & 10.12 (07)     & 0.69 (03) & 1.65 (18) \\
$I$     & 56688.1 (9)       &  9.75 (10)     & 0.62 (07) & 1.07 (21) \\
$J$     & $<$56687.2        & $<$9.36        & \nodata   & 0.26 (21) \\
$H$     & $\lesssim$56687.2 & $\lesssim$9.44 & \nodata   & 0.13 (16) \\
$K_{s}$ & $\lesssim$56687.2 & $\lesssim$9.20 & \nodata   & 0.00 (23) \\

\tablenotetext{a}{As determined relative to the \citet{Prieto06}
  templates for the optical bands and average of the methods listed by
  \citet{Matheson12} for the NIR bands.  Includes a 0.10~mag distance
  uncertainty to M82.}

\enddata

\end{deluxetable}

\begin{deluxetable}{lcr}
\tabletypesize{\footnotesize}
\tablewidth{0pt}
\tablecaption{Synthesised SN~2011fe UV Photometry\label{t:11fe_uv}}
\tablehead{
\colhead{MJD} &
\colhead{Filter} &
\colhead{Magnitude\tablenotemark{a}}}

\startdata

55801.174 & F218W & 17.35 \\
55804.270 & F218W & 15.51 \\
55807.431 & F218W & 14.02 \\
55811.423 & F218W & 13.37 \\
55814.440 & F218W & 13.20 \\
55817.677 & F218W & 13.11 \\
55823.630 & F218W & 14.08 \\
55835.269 & F218W & 15.46 \\
55841.320 & F218W & 15.85 \\
55855.203 & F218W & 16.39 \\
55801.174 & F225W & 16.86 \\
55804.270 & F225W & 14.69 \\
55807.431 & F225W & 13.24 \\
55811.423 & F225W & 12.73 \\
55814.440 & F225W & 12.65 \\
55817.677 & F225W & 12.74 \\
55823.630 & F225W & 13.57 \\
55835.269 & F225W & 14.95 \\
55841.320 & F225W & 15.35 \\
55855.203 & F225W & 15.95 \\
55801.174 & F275W & 14.62 \\
55804.270 & F275W & 12.28 \\
55807.431 & F275W & 11.05 \\
55811.423 & F275W & 10.73 \\
55814.440 & F275W & 10.82 \\
55817.677 & F275W & 11.18 \\
55823.630 & F275W & 12.16 \\
55835.269 & F275W & 13.71 \\
55841.320 & F275W & 14.19 \\
55855.203 & F275W & 14.83 \\
55801.174 & F336W & 12.52 \\
55804.270 & F336W & 10.49 \\
55807.431 & F336W &  9.55 \\
55811.423 & F336W &  9.25 \\
55814.440 & F336W &  9.33 \\
55817.677 & F336W &  9.70 \\
55823.630 & F336W & 10.63 \\
55835.269 & F336W & 12.37 \\
55841.320 & F336W & 12.95 \\
55855.203 & F336W & 13.64

\enddata

\tablenotetext{a}{Uncertainty in synthesised photometry is \about0.1~mag.}

\end{deluxetable}

\begin{deluxetable}{lrrrrrrrrrr}
\tabletypesize{\footnotesize}
\tablewidth{0pt}
\tablecaption{Spectral Phases (days relative to $B$-band maximum)\label{t:phase}}
\tablehead{
\colhead{SN~2014J} &
\colhead{$-6.4$} &
\colhead{$-4.6$} &
\colhead{$-2.5$} &
\colhead{$-0.4$} &
\colhead{$+2.7$} &
\colhead{$+6.5$} &
\colhead{$+8.4$} &
\colhead{$+11.3$} &
\colhead{$+14.4$} &
\colhead{$+24.1$}}

\startdata

SN~2009ig & \nodata & $-4.2$  & $-2.1$ & \nodata & \nodata & \nodata & $+8.5$ & \nodata & \nodata & \nodata \\
SN~2011fe & $-6.9$  & \nodata & $-3.0$ &  $0.0$  & $+3.2$  & \nodata & $+9.1$ & \nodata & \nodata & $+26.7$ \\
SN~2013dy & $-6.2$  & \nodata & $-2.1$ & $-0.4$  & \nodata & \nodata & $+8.8$ & $+12.4$ & $+14.4$ & $+21.2$

\enddata

\end{deluxetable}

\twocolumn

\label{lastpage}


\end{document}